\documentclass[%
reprint,
 showpacs,
 amsmath,amssymb,aps
]{revtex4-1}

\usepackage{graphicx}
\usepackage{dcolumn}
\usepackage{bm}

\usepackage[squaren]{SIunits}
\usepackage[export]{adjustbox}
\usepackage{xcolor}
\newcommand{\Berlinpro}{b{ERL}inPro}
\newcommand{\Htwo}{$\mathrm{H}_2^+$}
\newcommand{\CO}{$\mathrm{CO}^+$}

\newcommand{\CHfour}{$\mathrm{CH}_4^+$}

\newcommand{\Efield}{\mathbf{E}}

\usepackage{fancyhdr}
\begin{document}
\rhead{ \fancyplain{}{\textbf{\today}} }

\preprint{APS/123-QED}

\title{Investigation on the Ion Motion towards Clearing Electrodes in High-Brightness Electron Accelerators}
%
\author{Gisela P\"{o}plau\thanks{} }
\email{poeplau@math.uni-luebeck.de}
\affiliation{
Universit\"at zu L\"ubeck, L\"ubeck, Germany
}%

\author{Atoosa Meseck}
\email{atoosa.meseck@helmholtz-berlin.de}
\affiliation{%
 Helmholtz--Zentrum Berlin and Johannes Gutenberg-Universit\"at, Mainz, Germany\\
}%
\author{Fjodor Falkenstern}
\author{Michael Markert}
\author{Joachim Bartilla}
\affiliation{%
 Helmholtz--Zentrum Berlin, Germany\\
}%
\date{\today}
\begin{abstract}
High-brightness electron beams provided by modern accelerators require several measures to preserve their high quality and to avoid instabilities. The mitigation of the impact of residual ions is one of these measures. It is particularly important if high bunch charges in combination with high repetition rates are aimed for. This is because ions can be trapped in the strong negative electrical potential of the electron beam causing emittance blow-up, increased beam halo and longitudinal and transverse instabilities.

Over the last decades three ion-clearing strategies have been applied to counteract the degrading impact of the ions on the electron beam. These strategies are installation of clearing electrodes, introduction of gaps in filling pattern and beam shaking. Currently, their merit as clearing strategies for the next generation of high brightness accelerators such as energy recovery linacs (ERLs) are under intensive investigations by means of numerical and experimental studies. Of particular interest in this context is the performance of multi-purpose electrodes, which are designed such that they allow for a simultaneous ion-clearing and beam-position monitoring. Such electrodes will be installed in the \Berlinpro\ facility.
 
In this paper, we present numerical studies for the behavior of ions generated by electron bunches while passing through the field of clearing electrodes. The objective is to investigate the ion motion towards the electrodes and to study under which circumstances and up to which ratio an equilibrium between ion-generation and ion-clearing is established. Hereby, several ion species and configurations of electrodes are considered in combination with typical beam parameters of high-brightness electron accelerators. Furthermore, we present detailed numerical studies of the performance of multi-purpose clearing electrodes planned for \Berlinpro.

\end{abstract}

\maketitle


\section{\label{sec:intro}Introduction}
In an electron accelerator, ions generated from the residual gas molecules in the vacuum chamber can be trapped in the negative electrical potential of the beam and thus lead to an increase of the beam halo, emittance blow-up and transverse and longitudinal instabilities by interacting (oscillating) resonantly with the beam. Often the machine settings of electron accelerators providing high-brightness beams are such that all generated ions have an atomic mass larger than the critical mass (cp.~\cite{Hoffstaetter2006205,PhysRevSTAB.18.044401}) and are therefore trapped in the electrical potential of the bunch. Thus, the degradation of the beam quality caused by  ions is a serious threat to an optimum operation of modern electron accelerators and ion-clearing strategies are mandatory. There are several measures for mitigating the ion-trapping such as utilizing clearing electrodes, beam shaking or varying the fill pattern for example by using short or long gaps in bunch trains, also called clearing gaps. 
\par
Clearing electrodes seem to be simple and goal-oriented as they pull out the trapped ions from the vicinity of the beam. However, simulation studies of the clearing process by electrodes presented in~\cite{PhysRevSTAB.18.044401} have shown that the ions cannot be fully cleared out from the electron beam. A few ions remain always trapped. As the transitory bunches ionize the residual gas continuously, the question arises under which circumstances and up to which ratio, an equilibrium between ion generation and ion-clearing is established?
\par
Often in modern facilities, not just a single electrode, i.e.\ a biased stripe or button, but an electrode configuration consisting of several charged stripes or buttons positioned around the beam pipe at a given location is used to enable ion-clearing, see for example~\cite{Feikes_Pac2009,PhysRevAccelBeams.21.054401}. Please note that such a configuration of clearing electrodes is simply referred to as clearing electrode in this paper. 
\par
Clearing electrodes can be used to measure the beam position and clear simultaneously the ions from the vicinity of the beam, when the shapes and the positions of their biased stripes around the beam pipe are properly optimized for both tasks. The ERL facility \Berlinpro~\cite{berlinpro2018} currently under construction at Helmholtz--Zentrum Berlin, will utilize such multi-purpose clearing electrodes to clear ions and monitor the beam position. However in this case, the shapes and the positions of the single stripes have to satisfy simultaneously the constraints for ion-clearing and beam-position measurements. Thus, one might ask, how the multi-purpose clearing electrodes perform compared to classical clearing electrodes, where stripes or buttons with shapes optimized exclusively for ion-clearing are positioned as near as possible to the center of the beam pipe. 
\par
\begin{table}
  \centering
    \caption{Nominal parameters of a generic ERL and \Berlinpro.}
\begin{tabular}{cc}
  \hline
  \hline
  \multicolumn{2}{c}{Generic Bunch}\\
   \hline
  Nominal Beam Energy & \unit{80}{\mega\electronvolt} \\
  Nominal Beam Charge $Q$& \unit{77}{\pico\coulomb} \\
  Maximum Repetition Rate & \unit{1.3}{\giga\hertz} \\
  Transv. RMS Bunch Size $\sigma_t$& \unit{1.4}{\milli\metre} \\
  Bunch Length $\sigma_t$& \unit{2}{\pico\second} \\
  Vacuum Pressure & \unit{10^{-8}}{\milli\bbar}\\
  \hline
  \hline
    \multicolumn{2}{c}{\Berlinpro\ Bunch}\\
      \hline
  Nominal Beam Energy & \unit{50}{\mega\electronvolt} \\
  Nominal Beam Charge $Q$& \unit{77}{\pico\coulomb} \\
  Maximum Repetition Rate & \unit{1.3}{\giga\hertz} \\
  Transv. RMS Bunch Size $\sigma_t$& \unit{300}{\micro\metre} \\
  Bunch Length $\sigma_t$& \unit{2}{\pico\second} \\
  Vacuum Pressure & \unit{5\cdot10^{-10}}{\milli\bbar}\\
  \hline
  \hline
\end{tabular}
\label{beam}
\end{table}
In this paper, we first present a general study with four types of configurations with stripes and buttons commonly used as clearing electrodes~\cite{pc_joergfeikes,MasterEden}. For this, we use  bunch parameters for a generic energy recovery linac with parameters given in Table~\ref{beam}; we will call it a \emph{generic bunch} in the following. Furthermore we investigate the performance of a multi-purpose clearing electrode using the specific design chosen for the \Berlinpro\  facility together with the \Berlinpro\ beam parameters also shown in Table~\ref{beam}. Hereby we demonstrate the impact of different voltage configurations on the clearing performance. 
\par
For the presented numerical simulations we use the software tool CORMORAN developed by Compaec~e.G.~\cite{ipac2017}. CORMORAN tracks the ions taking into account the continuous generation of new ions according to given ionization rates. 
\par
The paper is organized as follows: a brief introduction of the tracking tool CORMORAN including the modeling of the ionization process as well as the ion-tracking is given in Sec.~II followed by a detailed description of simulations and results of the studies on equilibrium between ion generation and clearing for different electrode configurations in Sec.~III. The description and investigation on multi-purpose clearing electrodes are the subject of Sec.~IV followed by the discussion and concluding remarks in Sec.~V.
\section{\label{sec:02}Ion tracking with the tool CORMORAN}
\begin{figure*}[!tbh]
    \centering
 \includegraphics*[trim=15 0 110 0, clip,width=0.2284\textwidth]{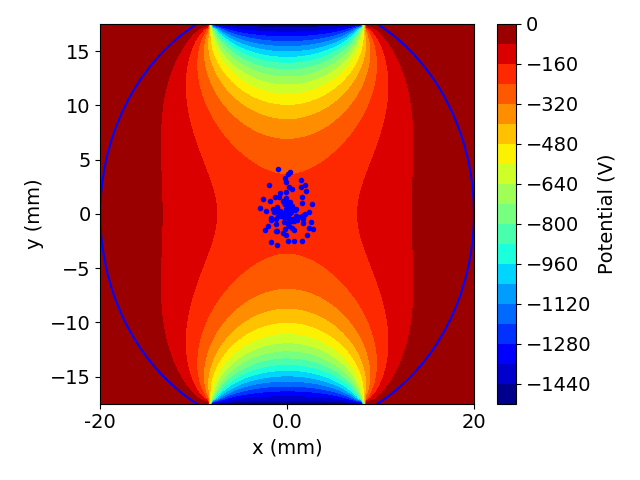}
 \includegraphics*[trim=15 0 110 0, clip,width=0.2284\textwidth]{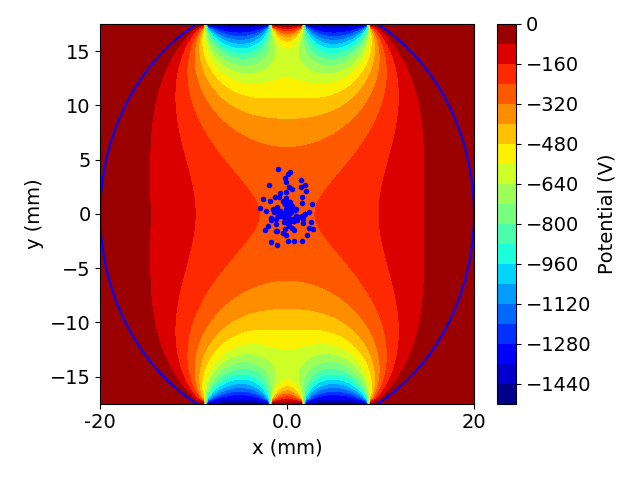}
\includegraphics*[trim=15 0 110 0, clip,width=0.2284\textwidth]{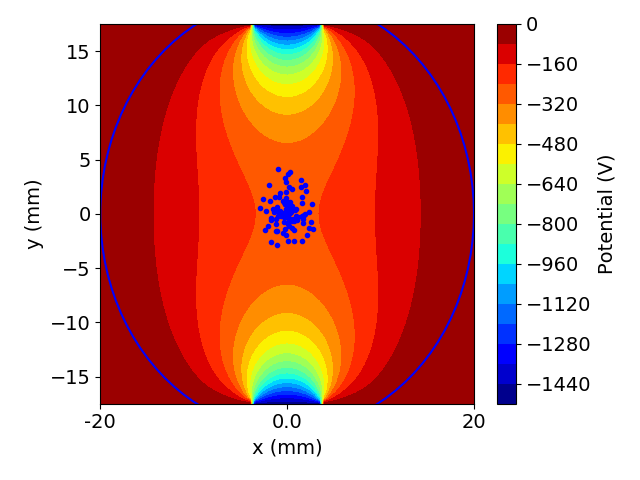}
 \includegraphics*[trim=15 0 20 0, clip,width=0.288\textwidth]{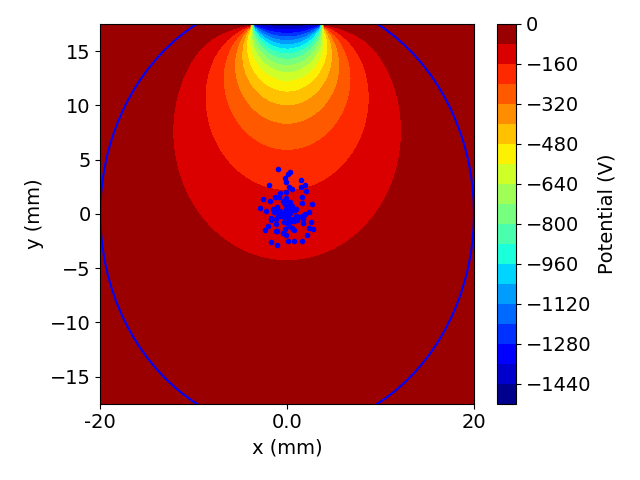}
    \caption{Potential of the electrodes~1 - 4 (from left to right), transversal cross-section shown with the first 100 generated ions.}
    \label{electrodes}
    \vspace*{-\baselineskip}
\end{figure*}
The software tool CORMORAN has been developed by Compaec~e.G., Rostock, in order to simulate the ionization process of residual gas and to further track the generated ions over long time ranges in interaction with passing bunches and under the influence of further external fields, such as the electrical field of clearing electrodes~\cite{ipac2017}. 
Hereby, secondary knock-out electrons generated with the ionization are not considered.
\par
Since only a few ions are moving simultaneously in the pipe, the interaction between ions are neglected. Hence, the forces   $\mathbf{F}_\mathrm{i} $ acting on the ions are the superposition of the forces caused by the field of the bunch $\Efield_\mathrm{b}$ and the field of the electrode $\Efield_\mathrm{elec}$:
\begin{equation}\label{forceions1}
        \mathbf{F}_\mathrm{i}  =
        q (\Efield_\mathrm{b}+\Efield_\mathrm{elec}),
\end{equation}
where $q$ denotes the charge of the ions.
\par
Furthermore, it is assumed, that the electrical fields of the high energy bunches do not change during the passage. 
Consequently, the  potential of the bunch and the clearing electrodes can be pre-computed. To do this Poisson's equation is solved with appropriate boundary conditions by means of the particle-mesh method (see~\cite{PhysRevSTAB.18.044401} and citations therein for more details), where Jacobi preconditioned conjugate gradients are applied as Poisson solver.
\par
Whereas the potential of the bunch is pre-computed at the beginning of the tracking process with CORMORAN, the potential of the electrodes is determined by means of the Python Poisson Solver developed by Compaec~e.G.~\cite{ipac2014} in order to enable a simple modification of the shape of the beam pipe and the location of the electrodes. The electrical field strength is computed from the potential values at the grid points of a tensor product mesh and is further interpolated to the position of the ions. For a more detailed description we refer to~\cite{PhysRevSTAB.18.044401}.
\subsection{Model of the Ionization Process}
In this paper two different mixtures of residual gas are considered (see Table~\ref{gasmixtures}). They were already applied for the simulations in~\cite{PhysRevSTAB.18.044401}. Both gases consist of \Htwo, \CHfour\ and \CO\ ions with mass numbers 2, 16 and 28, respectively. Gas A contains mainly the light \Htwo-ions, whereas the percentage of \Htwo\ in Gas~B amounts roughly to $50\%$. 
\par
According to~\cite{springerlink:10.1007/3-540-55250-2_41},
an ion of species $j$ is generated by a bunch passage at a rate of
\begin{equation}
R_{j} = c  N_{e} \sigma_{j}\frac{P}{k_{B}T} ,
\end{equation}
where $c$ denotes the speed of light, $N_{e} $ the number of electrons in the bunch, $\sigma_{j}$ the ionization cross-section of ion species $j$ (see~\cite{PhysRevA.6.1507}),
$P$ the vacuum pressure, $k_{B}$ the Boltzmann constant and $T$ the temperature. 
The resulting number of generic bunches that are necessary to generate a new ion per cm taking into account the percentage of the molecules in the gas mixture is given in Table~\ref{gasmixtures}. 
\par
The simulation procedure is built as follows: the first bunch passage generates one ion per cm, i.e.\  four ions are activated at a simulated length of \unit{4}{\centi\metre} (see also next section). Then according to Table~\ref{gasmixtures} four new \Htwo-ions are created for instance after each $37^{\mathrm{th}}$ bunch passage for gas mixture~A.
\par
For a more detailed picture we add the total number of generated ions after a time of \unit{50}{\micro\second} at a length of \unit{2}{\centi\metre} and \unit{4}{\centi\metre}, respectively.
These numbers are related to the simulation set-up of the generic case presented in Sec.~\ref{sec:03}.
\begin{table}[hbt]
   \centering
   \caption{Mixtures of ionized residual gas and the corresponding ionization process for the generic case.}
   \setlength{\tabcolsep}{3mm}
   \begin{tabular}{ccccc}
       \hline
        \textbf{ion }  &   & \textbf{ion gener.}&\multicolumn{2}{c}{\textbf{total no.\  of ions}}   \\
       \textbf{species}  &\textbf{ \%} & \textbf{after} &\multicolumn{2}{c}{{(\unit{50}{\micro\second})}}\\
                                 &                        &   \textbf{bunch no.} &\textbf{\unit{2}{\centi\metre}}&\textbf{\unit{4}{\centi\metre}}\\
       \hline
           Gas A        & & & &\\
       \hline
          \Htwo        & 98  &     37     & 3,514 & 7,028\\ 
	 \CHfour      & 1  &     565  & 232    &  464 \\ 
        \CO        	   & 1 &     660  & 198    &  396\\ 
       \hline
           Gas B       & & & &\\
       \hline
             \Htwo      &   48 &    75   &  1,734 &  3,468\\ 
             \CHfour   &  26 &     22  &  5,910 &11,820\\ 
             \CO         &  26 &     26  &  5,000 &10,000\\ 
      \hline
   \end{tabular}
   \label{gasmixtures}
\end{table}
\subsection{Model of Ion Cloud and Tracking}
The transverse positions of the successively generated ions correspond to a Gaussian distribution with the  same transverse rms size as the bunch, whereas longitudinally the ions are uniformly distributed over a distance of  \unit{2}{\centi\metre} and  \unit{4}{\centi\metre}, respectively. The ion velocities obey the Boltzmann distribution. 
The bunch passage is simulated with a time step of \unit{2}{\pico\second}. This leads to a ratio of \unit{2}{\pico\second} over \unit{84}{\pico\second} for \unit{2}{\centi\metre} and \unit{2}{\pico\second} over \unit{200}{\pico\second} for \unit{4}{\centi\metre} interaction region, respectively. After the interaction with a bunch, the ions are tracked further with 10 equally spaced time steps until the next bunch arrives. In total 65,000~bunch passages, i.e.\  \unit{50}{\micro\second} are simulated for the generic case (Sec.~\ref{sec:03}). The numerical studies for the multi-purpose electrode are performed with 1.3 million bunch passages and thus comprise \unit{10}{\milli\second} (Sec.~\ref{sec:04}). 
\par
The electrons within a bunch are modeled with 1~million macro-particles with a Gaussian distribution generated by the \emph{generator} procedure of ASTRA~\cite{ASTRA}. 
\par
Please note that for the investigation of the clearing behavior near neutralization the interaction between ions needs to be included in the simulation. As the presented studies are in a regime far from neutralization, we have refrained from it. 
Furthermore, for the investigation on the longitudinal ion motion toward electrodes the change in beam size along the accelerator has to be included in the simulation. While assuming a constant beam size over a length of a few centimeter seems reasonable, the change in beam size and the corresponding change of its attracting potential can be a major source of longitudinal motion towards the electrodes. Therefore, we state clearly that we have only investigated the clearing behavior of the electrodes and their direct vicinity,  as defined in the next section.  
\section{\label{sec:03}Investigation on the generic case}
In this section, we present numerical investigations on the clearing performance of four different electrode configurations. Thereby, we utilize the generic bunch described in Sec.~\ref{sec:intro}. In detail, we study the development and the level of the equilibrium between ion-clearing and ion-generation. In the simulation studies two cases are considered. In the first case the ions are generated within the field of the electrodes, i.e.\ here at a lengths of  \unit{2}{\centi\metre} longitudinally. In the second case, the ions are generated over a distance longer than the clearing electrodes  (\unit{4}{\centi\metre}), so that some of the ions start outside the electrodes allowing us to study the clearing performance in vicinity of the electrodes.  
\subsection{Four Types of Clearing Electrodes}
In the simulations the clearing electrodes are placed in the vacuum chamber. The shape of the vacuum chamber is modeled as a circular pipe with flattened parts at top and bottom for the clearing electrodes as shown in Fig.~\ref{electrodes}. The diameter in $x$-direction is \unit{40}{\milli\metre} and in $y$-direction \unit{34}{\milli\metre}. 
\par
\begin{figure}[]
   \centering
   \includegraphics*[width=65mm]{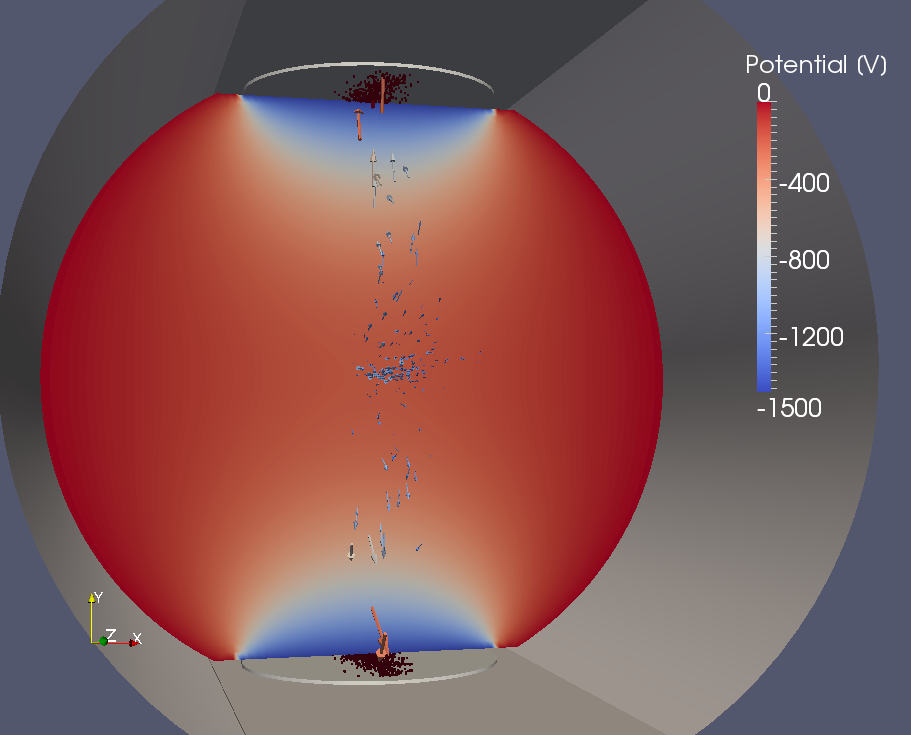}
   \includegraphics*[width=65mm]{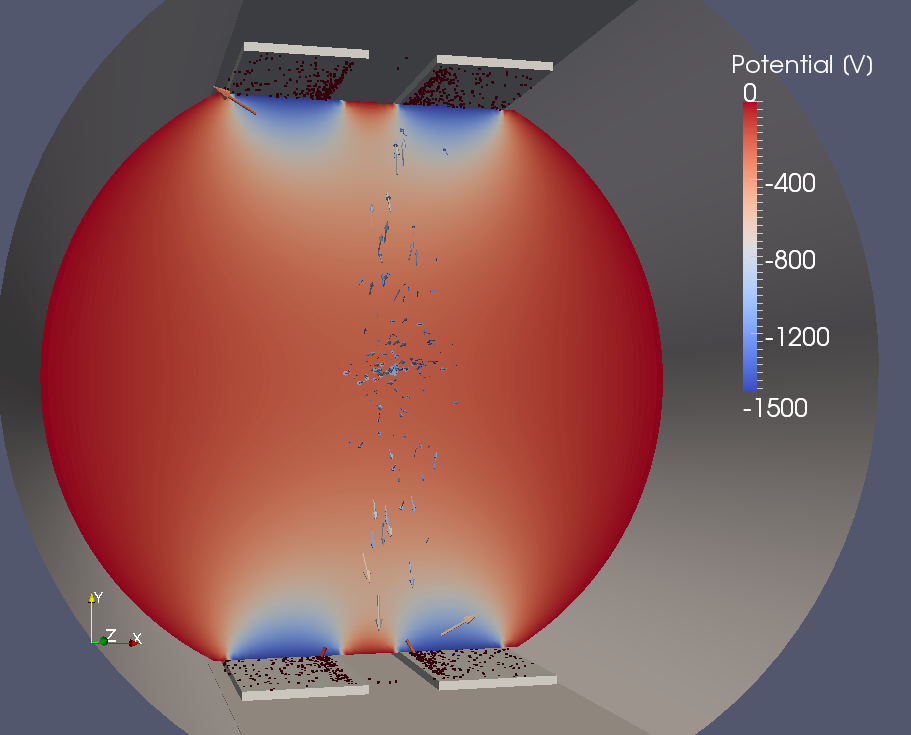}
   \caption{Distribution of the cleared ions at electrode~1 (top) and electrode~2 (bottom) with the clearing situation after \unit{50}{\micro\second}. The ions are generated over a length of \unit{2}{\centi\metre} within the field of the clearing electrodes.}
   \label{ions-at-elec1-2-trans}
\end{figure}

The following four types of electrodes are investigated:
\begin{itemize}
\item \emph{electrode~1} consists of two round electrode-plates, i.e.\ buttons, with a diameter of \unit{16}{\milli\metre} positioned vertically opposite to each other at top and bottom of the beam pipe;
\item \emph{electrode~2} has four rectangular electrode-plates, i.e.\ stripes, with a width of \unit{7}{\milli\metre} and a length of \unit{20}{\milli\metre}, placed pairwise in parallel with a distance of \unit{4}{\milli\metre} whereby the two pairs are positioned vertically opposite to each other;
\item  \emph{electrode~3} consists of two stripe electrodes each of a width of \unit{8}{\milli\metre} and length of \unit{60}{\milli\metre} positioned vertically opposite to each other;
\item  \emph{electrode~4} is a single stripe electrode of the same shape as the electrode~3 but placed at top of the beam pipe.
\end{itemize}
The potentials of the first three electrodes are symmetric with respect to the $y$-axis. Fig.~\ref{electrodes} shows the potential of all four electrodes at a voltage of \unit{-1500}{V}. The design of electrode~1 and electrode~2 is depicted in Fig.~\ref{ions-at-elec1-2-trans} in more detail.
\par
Please note that the symmetrical potentials of top and bottom plates of the electrodes~1--3 lead to an electrical field that cancels  in the center of beam pipe. This diminishes the undesired impact of the electrodes on the low energy electron bunches. 
The electrode 4 however has still a certain field strength in the center of the pipe which on the one hand enhances its efficiency and on the other hand could affect a low energy electron beam. 
\par
\begin{figure}[]
    \centering
\includegraphics*[width=0.238\textwidth]{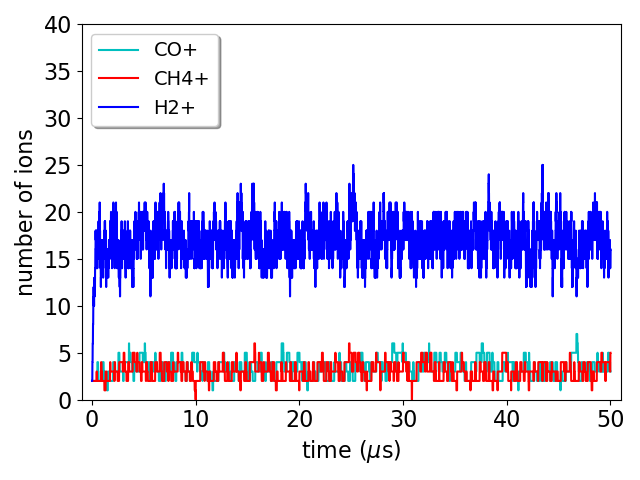}
\includegraphics*[width=0.238\textwidth]{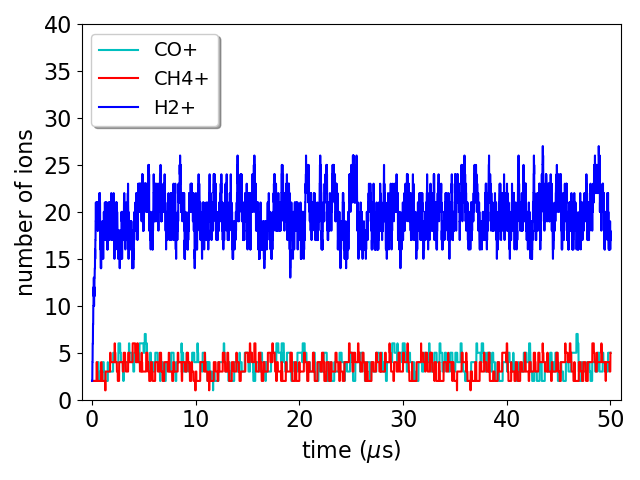}
\includegraphics*[width=0.238\textwidth]{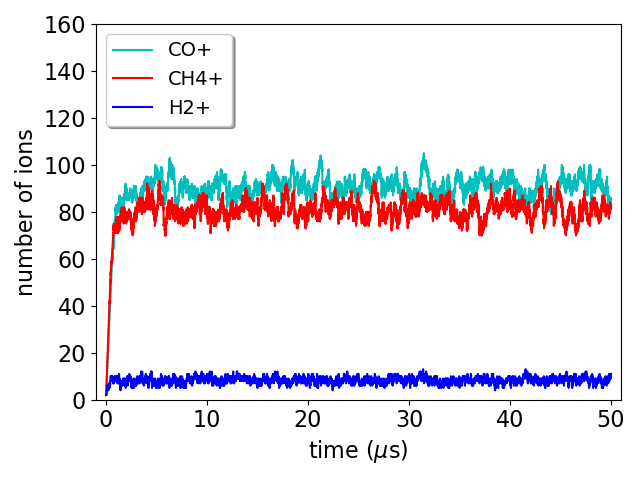}
\includegraphics*[width=0.238\textwidth]{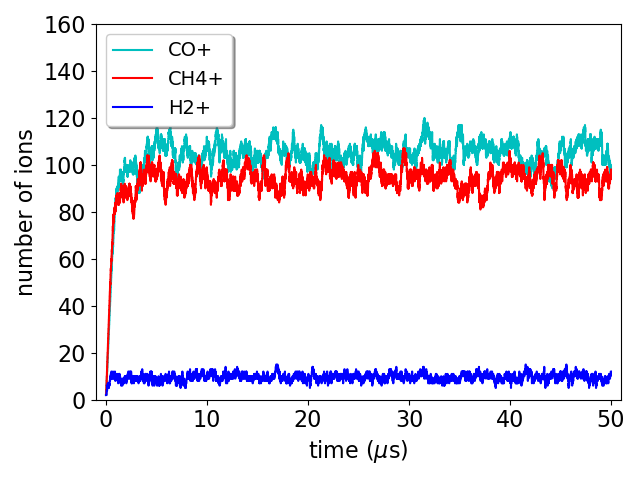}
\caption{Comparison of electrode 1 (left) and 2 (right) at a voltage of \unit{-1500}{\volt}:  the number of remaining ions is shown as a function of clearing time for gas mixture  A (top) and B (bottom). The ions are generated over a length of \unit{2}{\centi\metre} within the field of the clearing electrodes. The total number of generated ions is given in Table~\ref{gasmixtures}.}
    \label{electrods12_1500_gasB}
\end{figure}

\begin{figure}[]
    \centering
\includegraphics*[width=0.238\textwidth]{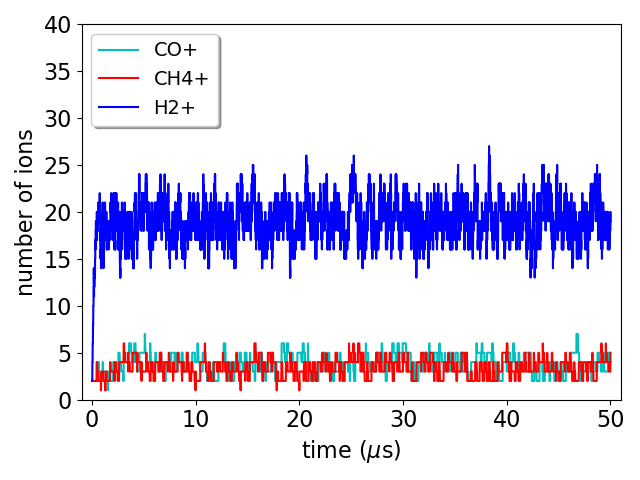}
\includegraphics*[width=0.238\textwidth]{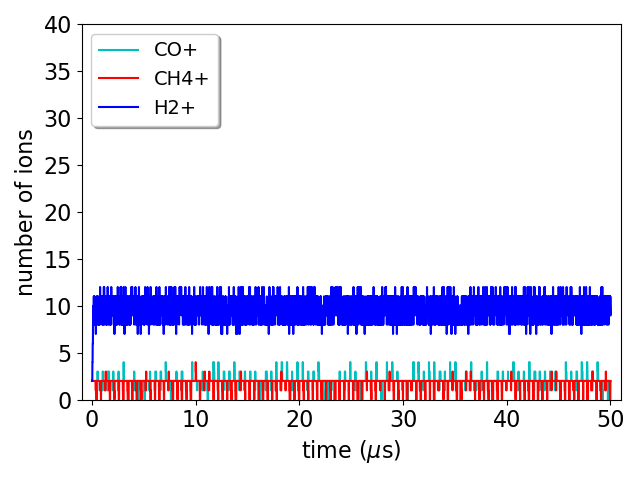}
\includegraphics*[width=0.238\textwidth]{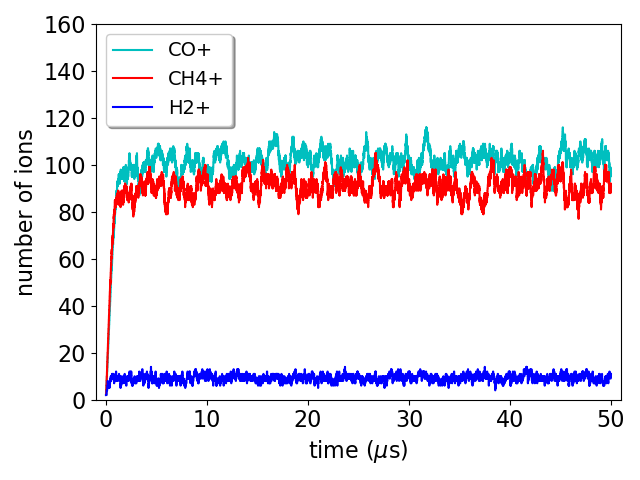}
\includegraphics*[width=0.238\textwidth]{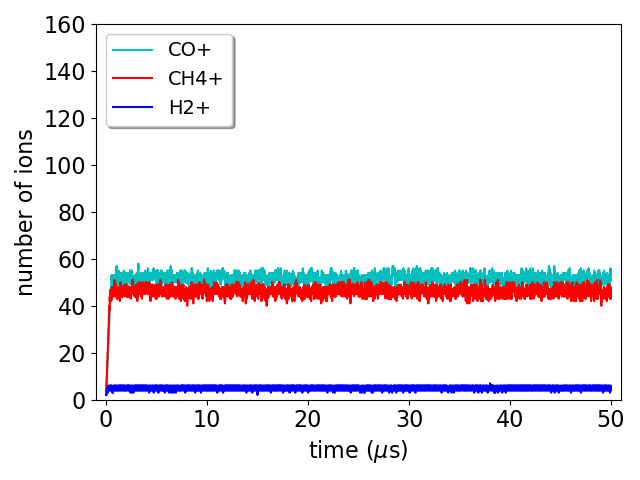}
\caption{Comparison of electrode 3 (left) and 4 (right) at a voltage of \unit{-1500}{\volt}:  the number of remaining ions is shown as a function of clearing time for gas mixture  A (top) and B (bottom). The ions are generated over a length of \unit{2}{\centi\metre} within the field of the clearing electrodes. The total number of generated ions is given in Table~\ref{gasmixtures}.}
    \label{electrods34_1500_gasB}
\end{figure}

\begin{figure}[]
    \centering
\includegraphics*[width=0.238\textwidth]{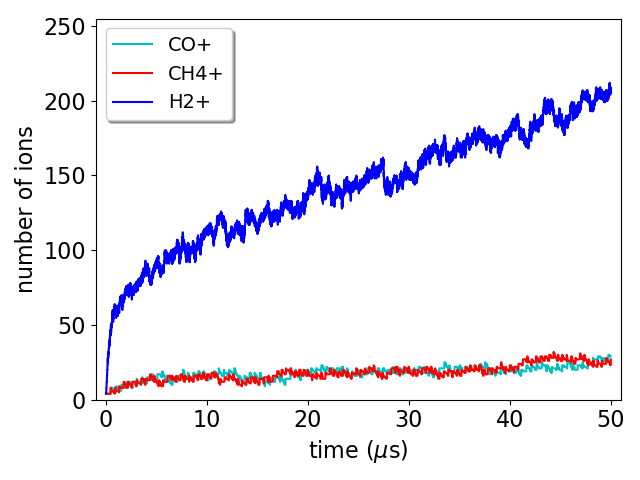}
\includegraphics*[width=0.238\textwidth]{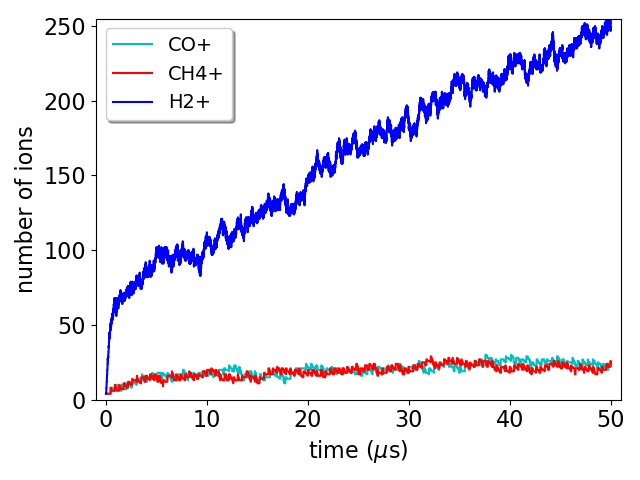}
\includegraphics*[width=0.238\textwidth]{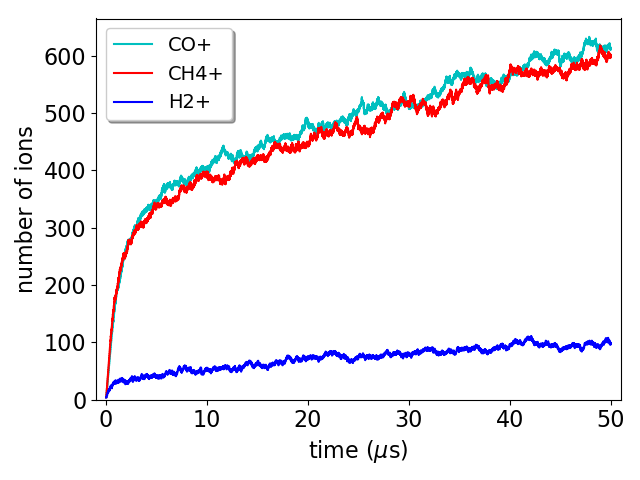}
\includegraphics*[width=0.238\textwidth]{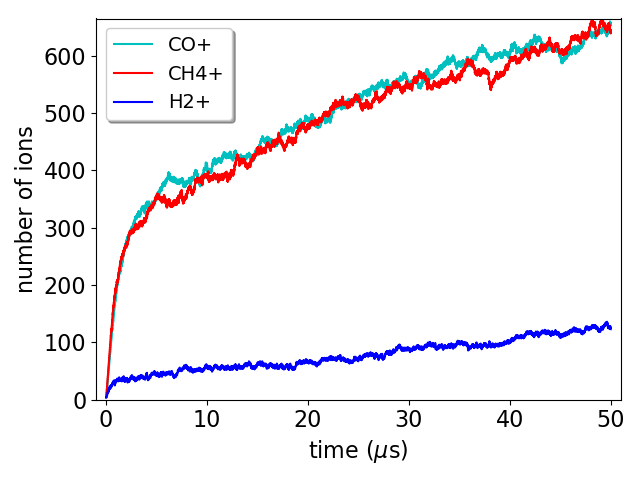}
\caption{Comparison of electrode 1 (left) and 2 (right) at  \unit{-1500}{\volt}:  the number of remaining ions is shown as a function of clearing time for gas mixture  A (top) and B (bottom). The ions are generated over a length of \unit{4}{\centi\metre} within the field of the clearing electrodes. The total number of generated ions is given in Table~\ref{gasmixtures}.}
    \label{electrods12_1500}
\end{figure}

\begin{figure}[]
    \centering
 \includegraphics*[width=0.238\textwidth]{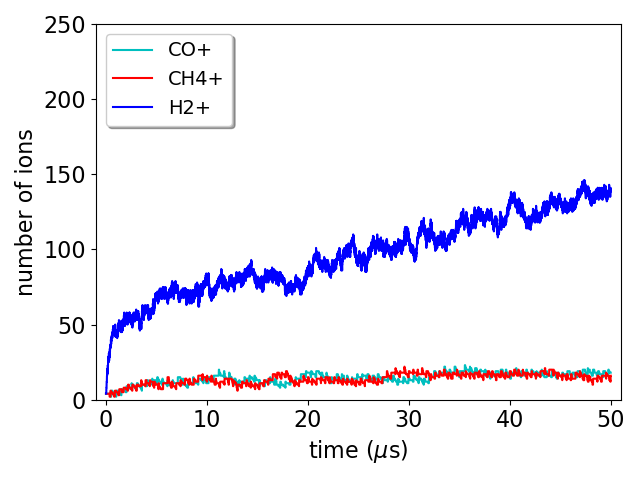}
 \includegraphics*[width=0.238\textwidth]{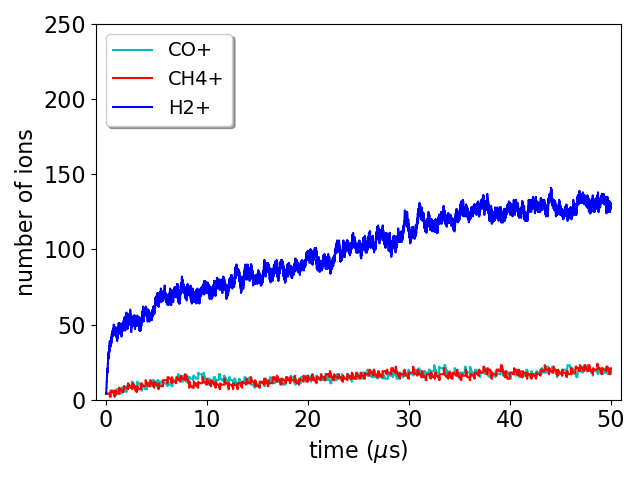}
    \centering
 \includegraphics*[width=0.238\textwidth]{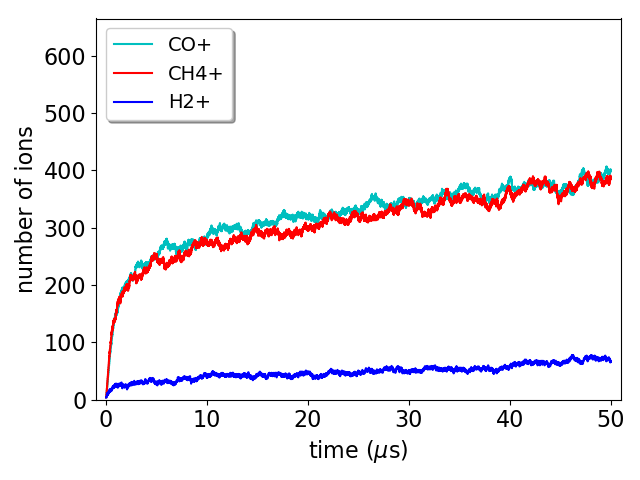}
 \includegraphics*[width=0.238\textwidth]{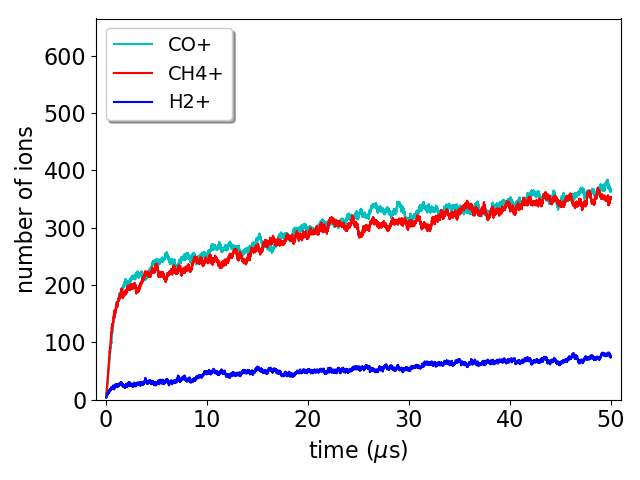}
\caption{Comparison of electrode 1 (left) and 2 (right) at a voltage of \unit{-2700}{\volt}:  the number of remaining ions is shown as a function of clearing time for gas mixture  A (top) and B (bottom). The ions are generated over a length of \unit{4}{\centi\metre}. The total number of generated ions is given in Table~\ref{gasmixtures}.}
    \label{electrods12_2700}
\end{figure}

\begin{figure}[]
    \centering
 \includegraphics*[width=0.238\textwidth]{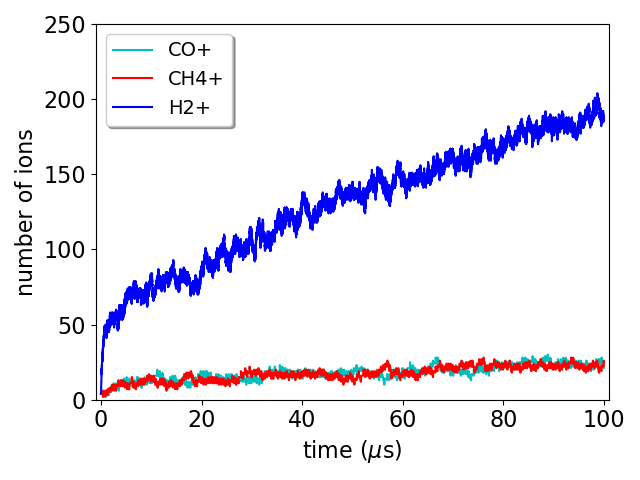}
 \includegraphics*[width=0.238\textwidth]{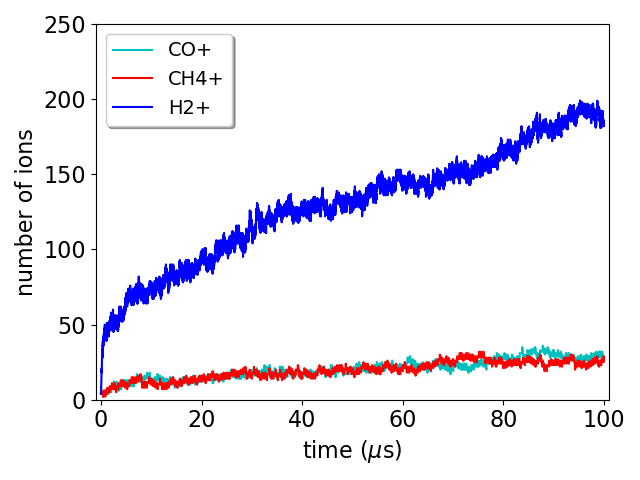}
 \includegraphics*[width=0.238\textwidth]{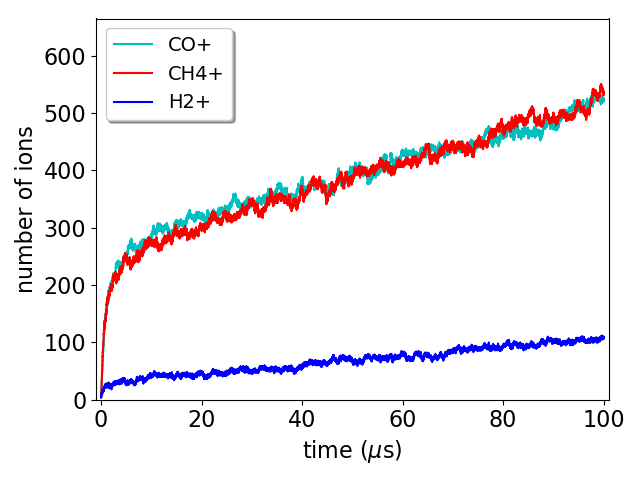}
 \includegraphics*[width=0.238\textwidth]{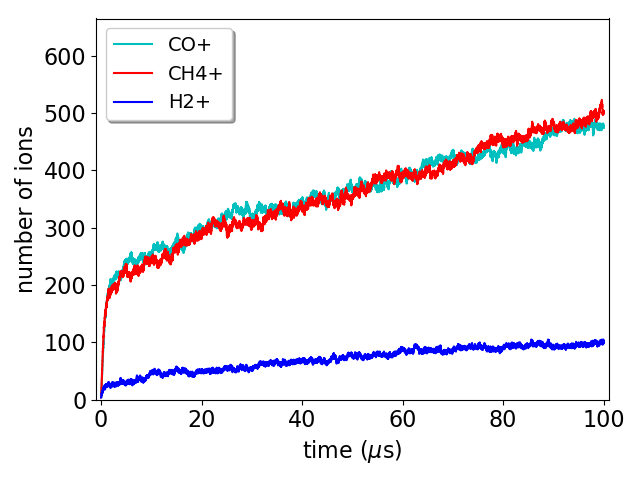}
\caption{Comparison of electrode 1 (left) and 2 (right) at  a voltage of \unit{-2700}{\volt}:  the number of remaining ions is shown as a function of clearing time for gas mixture  A (top) and B (bottom). The ions are generated over a length of \unit{4}{\centi\metre}. The total number of generated ions is a factor 2 higher than the numbers given in Table~\ref{gasmixtures}.}
    \label{electrods12_100mum}
\end{figure}

\begin{figure}[]
    \centering
 \includegraphics*[width=0.238\textwidth]{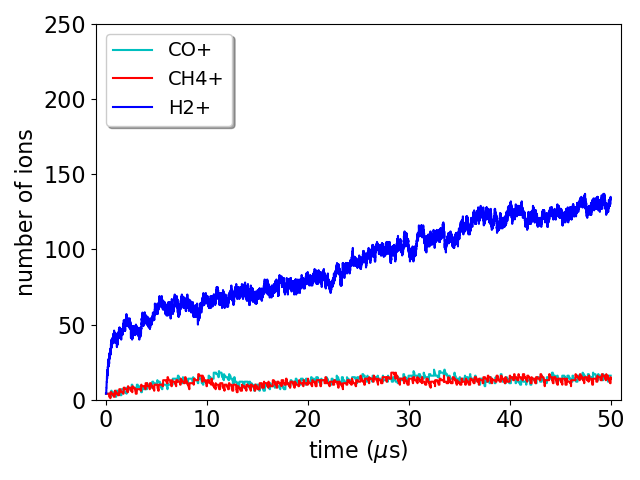}
 \includegraphics*[width=0.238\textwidth]{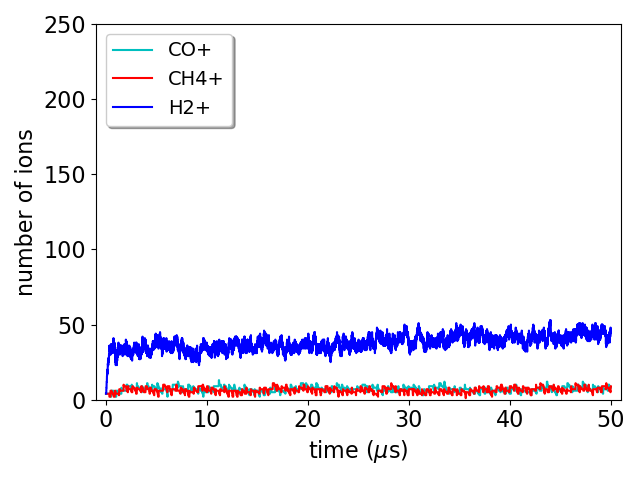}
 \includegraphics*[width=0.238\textwidth]{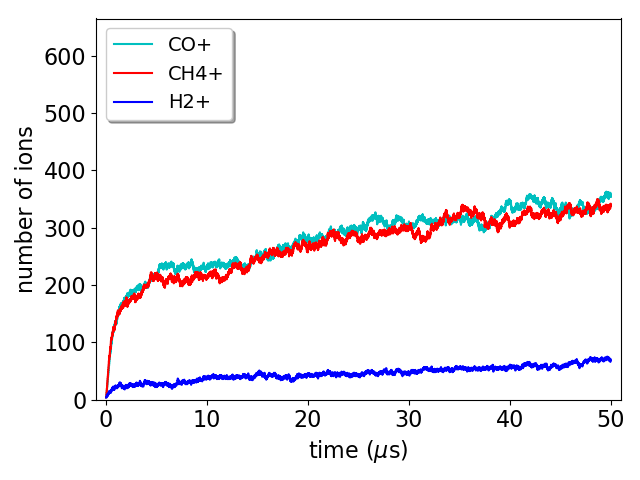}
 \includegraphics*[width=0.238\textwidth]{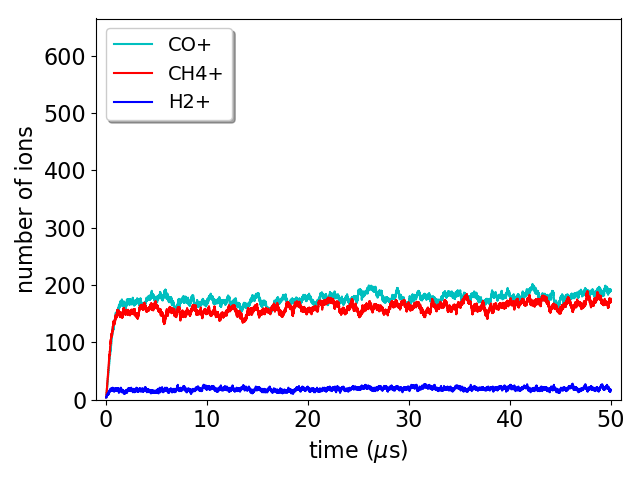}
\caption{Comparison of electrode 1 (left) and 2 (right) at  a voltage of \unit{-3500}{\volt}:  the number of remaining ions is shown as a function of clearing time for gas mixture  A (top) and B (bottom). The ions are generated over a length of \unit{4}{\centi\metre}. }
    \label{electrods12_100mum2}
\end{figure}

\begin{figure}[]
    \centering
 \includegraphics*[width=0.238\textwidth]{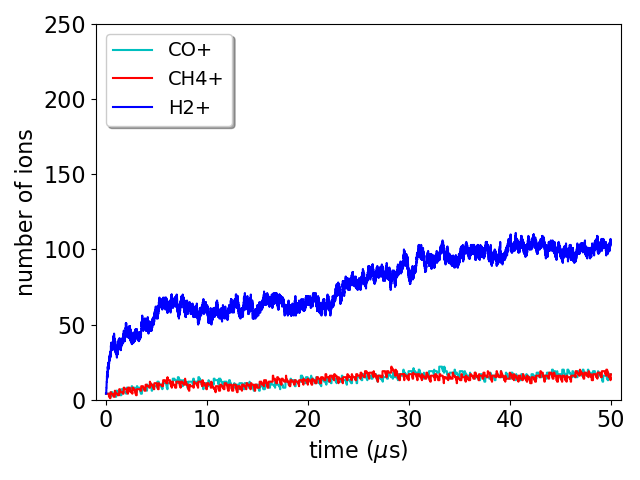}
 \includegraphics*[width=0.238\textwidth]{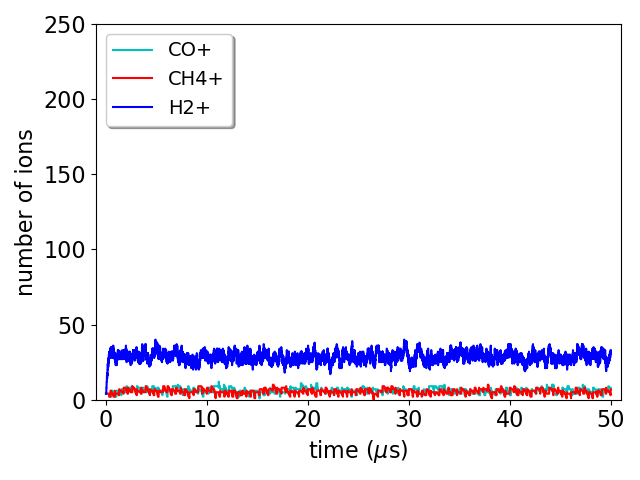}
 \includegraphics*[width=0.238\textwidth]{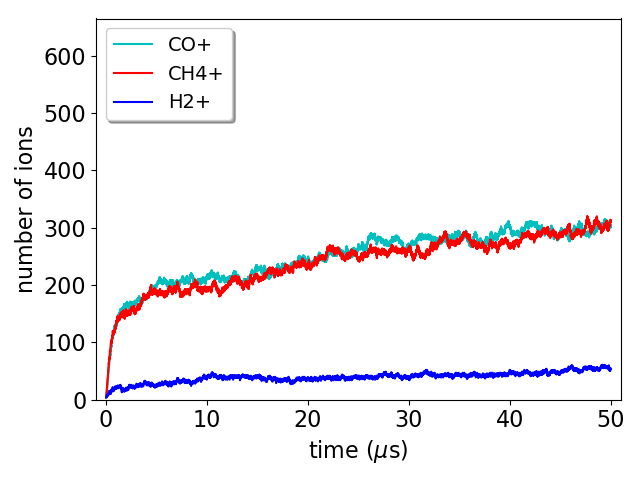}
 \includegraphics*[width=0.238\textwidth]{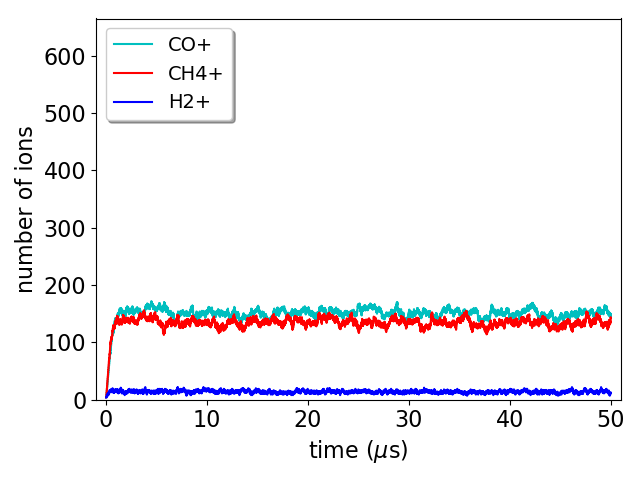}
\caption{Comparison of electrode 1 (left) and 2 (right) at a voltage of \unit{-4000}{\volt}:  the number of remaining ions is shown as a function of clearing time for gas mixture  A (top) and B (bottom). The ions are generated over a length of \unit{4}{\centi\metre}. }
    \label{electrods12_4000V}
\end{figure}

\subsection{Numerical Results}
We begin by examining the behavior of ions generated around the pipe centre within the volume occupied by the beam over a length of \unit{2}{\centi\metre}, i.e.\ within the field of clearing electrodes for all four investigated electrode-configurations. A voltage of \unit{-1500}{V} is applied to all electrodes and the development and the level of the equilibrium between ion-clearing and ion-generation for both gas mixtures is studied. Figs.~\ref{electrods12_1500_gasB} and \ref{electrods34_1500_gasB} show the comparison of the number of trapped ions of the same species after \unit{50}{\micro\second}. For all four types of clearing electrodes equilibrium between ion-generation and ion-clearing is achieved after a short rising time of less than \unit{1}{\micro\second}. The level of trapped ions is in all cases less than $10\%$ of the total number of generated ions (see Table~\ref{gasmixtures}) and quite similar for the electrodes~1 -- 3 with a small advantage towards electrode~1. Having still some measurable field on the pipe axis, electrode~4 shows the most efficient clearing. Its equilibrium level is only about a half of the other three electrodes. However, the residual field on the axis, which is still about \unit{- 200}{\volt}, causes a deflection of \unit{5}{ \micro\radian} of the generic beam.
\par
Fig.~\ref{ions-at-elec1-2-trans} represents the final distribution of all ions cleared within \unit{50}{\micro\second} together with location and velocity of the ions still remaining in the beam pipe for clearing electrodes~1 and 2, if a voltage of \unit{-1500}{\volt} is applied. These figures also show where the cleared ions hit the stripes of the clearing electrodes.
\par
Next we study the clearing performance in the vicinity of the longitudinal boundaries of the locally restricted electrodes~1 and 2. For this, ions are generated longitudinally over a distance of \unit{4}{\centi\metre}. Consequently, some of the ions are not situated within the attracting field of the clearing electrode but get trapped by the field of the bunch. Figs.~\ref{electrods12_1500} and~\ref{electrods12_2700} show comparisons of the number of trapped ions for the electrodes 1 and 2 for both gas mixtures and two different clearing voltages: \unit{-1500}{\volt} and  \unit{-2700}{\volt}, respectively. Obviously, in these cases the ions accumulate even with a clearing voltage as high as \unit{-2700}{\volt}. Of course the accumulation rates are higher for lower clearing voltages. 
\par
Comparing Figs.~\ref{electrods12_1500} and \ref{electrods12_2700}, one can conclude that the clearing with electrode~1 is slightly more efficient at the lower voltage for gas mixture A. However, for a clearing voltage of \unit{-2700}{\volt} clearly less accumulation can be observed for the heavier ions  \CHfour\ and \CO with electrode~2. The special composition of electrode~2 which allows for higher fields around the pipe center seems to be beneficial to the clearing process of heavier ions. This is of interest because generally heavier ions tend to resist clearing attempts~\cite{icap2015}. In order to better understand the ion accumulation in this case, we double the simulation time for the \unit{-2700}{\volt}. Fig.~\ref{electrods12_100mum} shows the result of these simulations. After \unit{100}{\micro\second} the rate of accumulation for both electrodes seems similar, although one might see a hint of the onset of accumulation stop for electrode 2.
\par
However, a simple possibility to stop the ion accumulation and clear also the vicinity of the clearing electrodes is to increase the voltage. Fig.~\ref{electrods12_100mum2} shows a comparison of the trapped ions for  electrodes~1 and 2 for both gas mixtures with a clearing voltage of \unit{-3500}{\volt}. The clearing electrode~2 is clearly more efficient in this case as it almost stops the ion accumulation while the clearing electrode~1 allows an increase of the number of trapped ions. Although increasing the clearing voltage to \unit{-4000}{\volt} improves the clearing performance of electrode 1 as shown in Fig.~\ref{electrods12_4000V}, a clear stop of accumulation seems to require even a higher voltage. In summary, the clearing electrodes 1 and 2 show a very good clearing behavior within their clearing fields as shown in Fig.~\ref{electrods12_1500_gasB}, but significantly differ in clearing performance in their vicinity, where the electrode~2 is more efficient.
\section{\label{sec:04}Investigation on the Multi-purpose electrode for \Berlinpro}
The motivation for the development of multi-purpose clearing electrodes is the lack of space for electrodes exclusively devoted to ion-clearing at the \Berlinpro\ machine. The needed space can be significantly reduced when combing the detection of the beam position with the extraction of parasitic ions by using the same BPM stripes for both tasks.  
\begin{figure}[]
  \centering
   \includegraphics*[width=70mm]{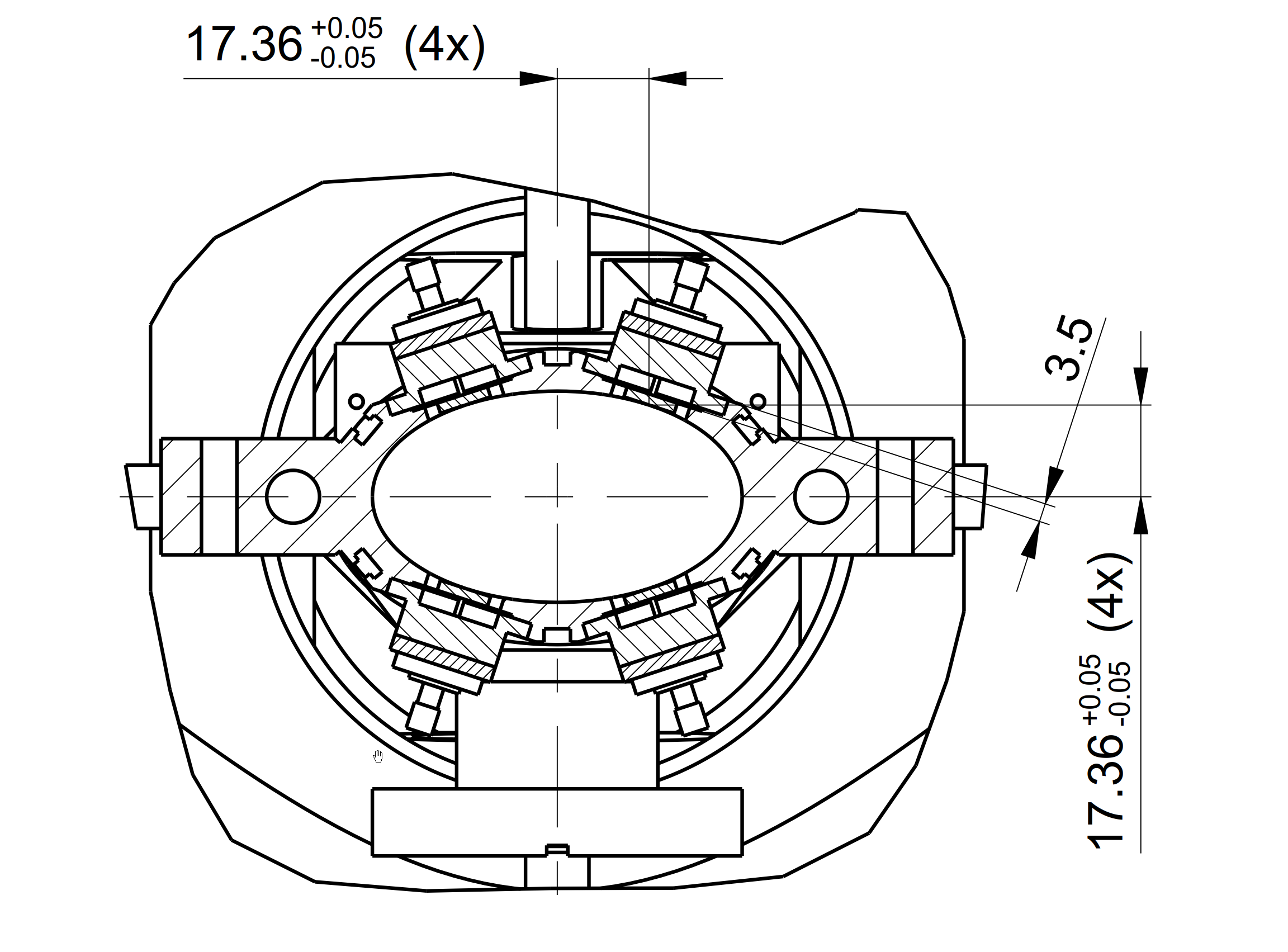}\\[3ex]
   \includegraphics*[width=65mm]{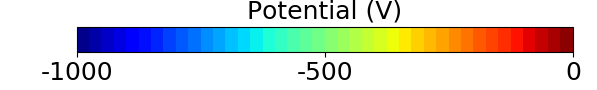}\\[1ex]
   \includegraphics*[width=65mm]{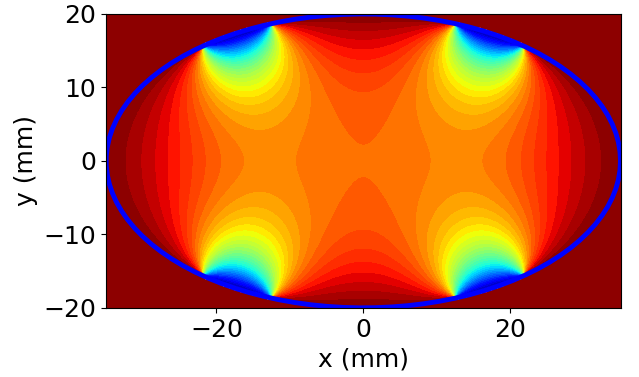}
   \caption{Technical drawing of the \Berlinpro\ multi-purpose electrode (top) with a plot of the potential for voltage-configuration~1. }
   \label{Atibild}
\end{figure}

\subsection{The Design of the \Berlinpro\ Multi-Purpose Electrode}
\begin{figure}[]
   \centering
   \includegraphics*[width=65mm]{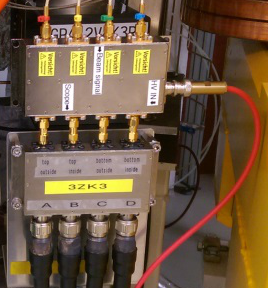}
   \caption{Installation of an ion-clearing box at MLS (PTB). On the right side, there is the connector for the high voltage (red cable);
on top of the box, it goes to the pick-up electrodes (signal input), on the bottom of the box, to the evaluation electronics.}
   \label{clearingbox}
\end{figure}

The multi-purpose clearing electrodes consist of four rectangular electrode-plates, i.e.\ stripes, with a width of \unit{10}{\milli\metre} and a length of \unit{38}{\milli\metre}, placed pairwise in parallel with a distance of \unit{24.72}{\milli\metre}, whereby the two pairs are positioned vertically opposite to each other. Fig.~\ref{Atibild} shows a technical drawing of the electrodes and the corresponding model used in the presented simulation studies.  The larger distance of \unit{24.72}{\milli\metre} between parallel plates is needed for an optimum beam-position detection. As shown in Fig.~\ref{clearingbox} with the multi-purpose electrodes also an ``ion-clearing'' box (top) has to be installed in addition to the BPM box (bottom). This box is simply inserted between the BPM box and stripe lines (SMA connectors). 

\par
\begin{figure}[]
   \centering
   \includegraphics*[width=80mm]{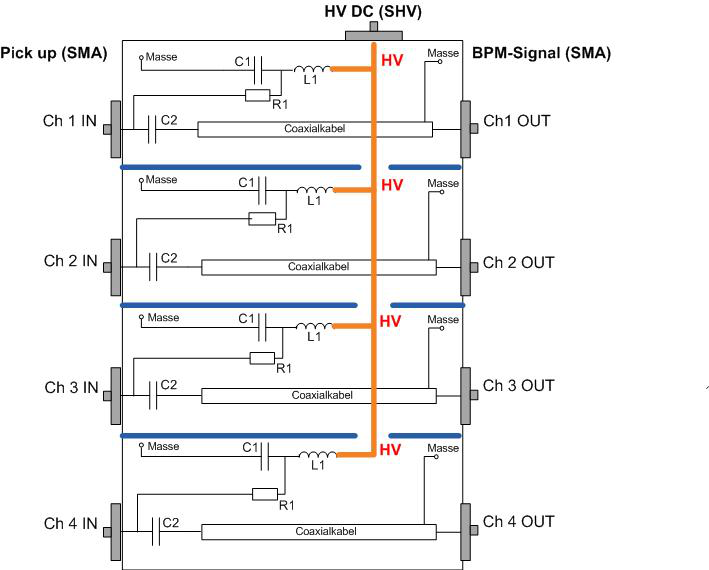}
   \caption{Schematic circuit of the ion-clearing box.}
   \label{circuit}
\end{figure}

The ion-clearing box allows to supply the pick-up electrodes with high voltage, while at same time detecting the signal for beam position from the same pick-up electrodes. It has been developed and successfully tested in the laboratory at HZB with the schematic circuit depicted in Fig.~\ref{circuit}.
Here, the coil L1 and the resistor R1 represent a low-pass filter which only allows DC voltage (HV) to pass through. The high voltage goes directly to the pick-up electrode.  Resistor R1 is also used to decouple the HF. The capacitor C2, on the other hand, should only ``pass''
 the high-frequency pick-up signal generated by the beam - but block the high voltage. The capacitor C1 serves as a smoothing capacitor. %

\par
\begin{figure}[h]
   \centering
   \includegraphics*[width=35mm]{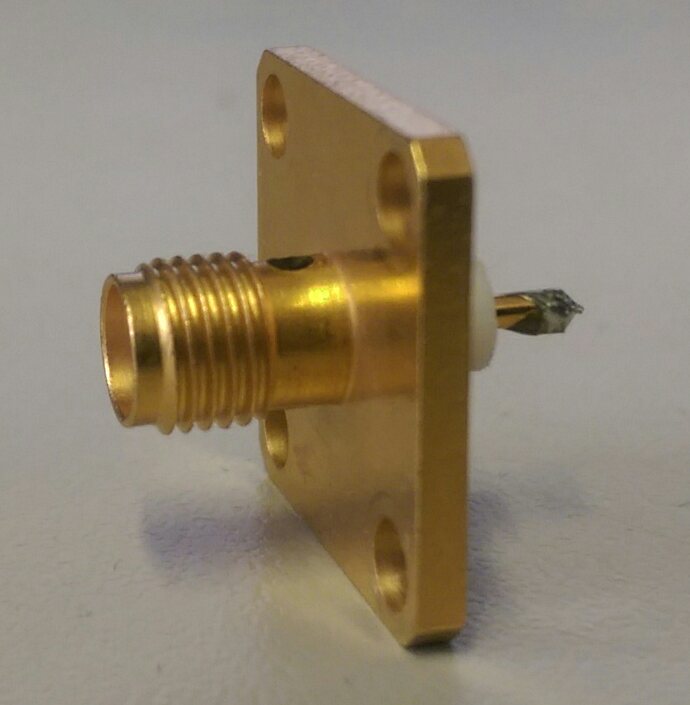}
   \includegraphics*[width=35mm]{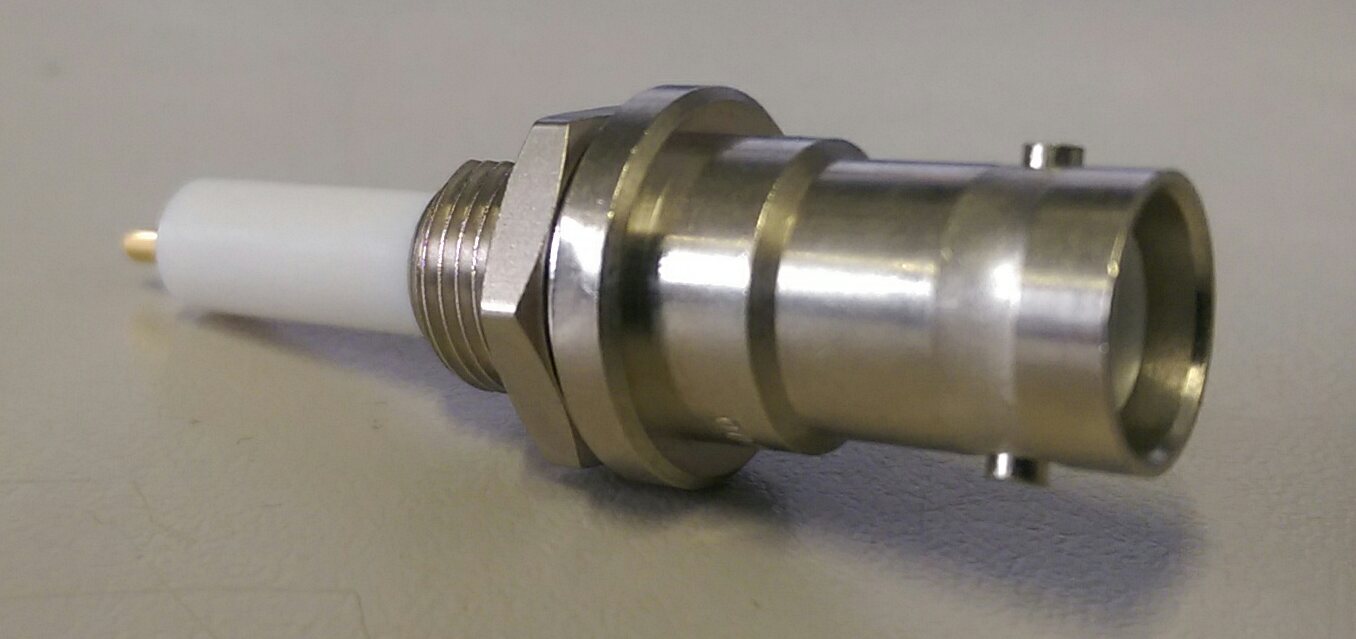}
   \caption{SMA connector (left) and SHV  connector (right).}
   \label{connector}
\end{figure}

The SMA connectors (Fig.~\ref{connector}, left) are voltage-resistant up to \unit{1}{\kilo\volt}, so higher voltages must not be applied under any circumstances. Long-term tests with \unit{1}{\kilo\volt} were also successfully carried out in the laboratory. 
The SHV connectors (Fig.~\ref{connector}, right) are less critical because they guarantee a dielectric strength of up to \unit{5}{\kilo\volt}. The electronic components used within the ion-clearing box were selected such that they can be permanently loaded with \unit{1}{\kilo\volt}.

\begin{figure}[]
   \centering
   \includegraphics*[width=65mm]{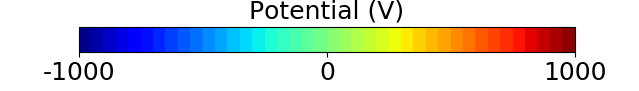}
    \includegraphics*[width=42mm]{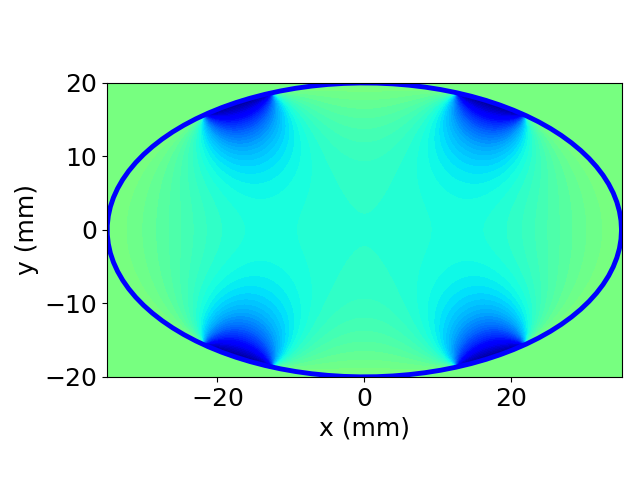}
   \includegraphics*[width=42mm]{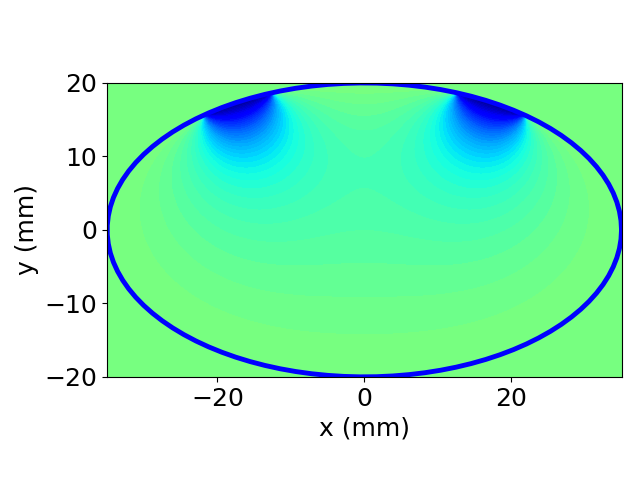}
   \includegraphics*[width=42mm]{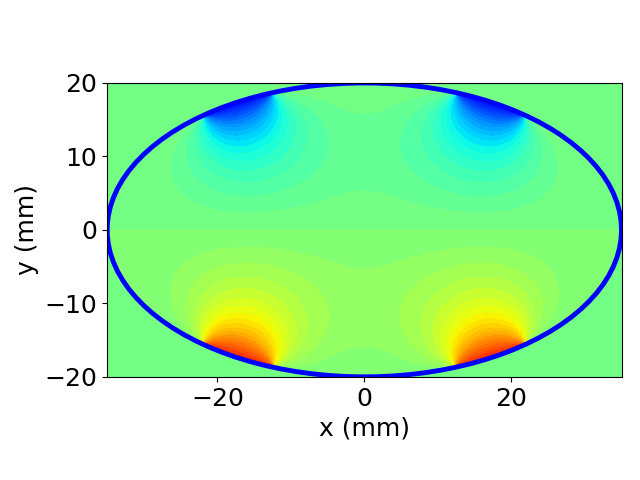}
   \includegraphics*[width=42mm]{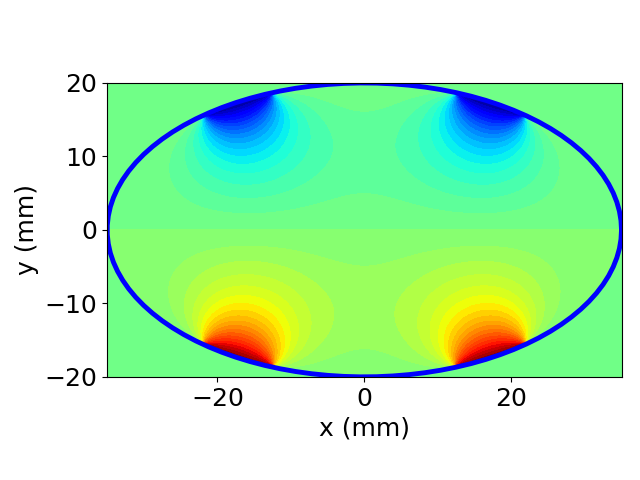}
   \caption{Potential of the \Berlinpro\  multi-purpose electrode in transversal cross-section: voltage-configuration~1 (top, left), voltage-configuration~2 (top, right), voltage-configuration~3 (bottom, left) and voltage-configuration~4 (bottom, right).}
   \label{bpro_potential}
\end{figure}

\subsection{Numerical Results}
The vacuum pressure in the case of the \Berlinpro\ facility is about \unit{5\cdot10^{-10}}{\milli\bbar} and thus much lower than in the generic case with \unit{10^{-8}}{\milli\bbar}. Therefore, the time scales for ion accumulation or reaching an equilibrium are about \unit{10}{\milli\second} for simulations presented in this section and thus much longer than the cases presented in the previous section. The number of generated ions in \unit{10}{\milli\second} for gas mixtures A and B are summarized in Table~\ref{gasmixturesbpro}. 
\begin{table}[hbt]
   \centering
   \caption{Mixtures of ionized residual gas and the corresponding ionization process for the \Berlinpro\ case.}
   \setlength{\tabcolsep}{3mm}
   \begin{tabular}{cccc}
       \hline
        \textbf{ion }  &   & \textbf{ion gener.}& \textbf{total no.\  of ions}  \\
       \textbf{species}  &\textbf{ \%} & \textbf{after} & (\unit{10}{\milli\second})\\
                                 &                        &   \textbf{bunch no.} &\textbf{\unit{4}{\centi\metre}}\\
       \hline
           Gas A        & & &  \\
       \hline
          \Htwo        & 98  & 18,500     & 2,812  \\ 
	 \CHfour      & 1   &  282,500   & 188     \\ 
        \CO        	   & 1  &   330,000   & 160     \\ 
       \hline
           Gas B       & & &  \\
       \hline
             \Htwo      &   48 &  37,000     &  1,388  \\ 
             \CHfour   &   26 &  11,000     &  4,728  \\ 
             \CO         &   26 &   13,000    &  4,000  \\ 
      \hline
   \end{tabular}
   \label{gasmixturesbpro}
\end{table}

\par
As discussed above, the maximum possible voltage for the \Berlinpro\ multi-purpose clearing electrodes amounts to \unit{\pm 1000}{\volt}. To examine the development and the level of the equilibrium between ion-clearing and ion-generation in the \Berlinpro\ case numerical simulations are performed with the following voltages:
\begin{itemize}
\item \emph{voltage-configuration~1}: \unit{-1000}{\volt} supplied to all four stripes of the clearing electrode;
\item \emph{voltage-configuration~2}: \unit{-1000}{\volt} supplied to the two upper stripes and \unit{0}{\volt} supplied to the two lower stripes of the clearing electrode;
\item \emph{voltage-configuration~3}: \unit{-750}{\volt} supplied to the two upper stripes and \unit{750}{\volt} supplied to the two lower stripes of the clearing electrode;
\item \emph{voltage-configuration~4}: \unit{-1000}{\volt} supplied to the two upper stripes and \unit{1000}{\volt} supplied to the two lower stripes of the clearing electrode.
\end{itemize}
Fig.~\ref{bpro_potential} shows the corresponding potentials in transversal cross-section.
\par
\begin{figure}[]
   \centering
   \includegraphics*[width=0.238\textwidth]{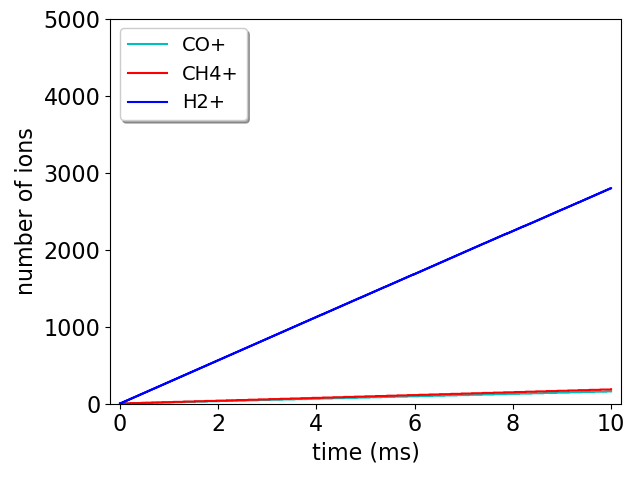}
   \includegraphics*[width=0.238\textwidth]{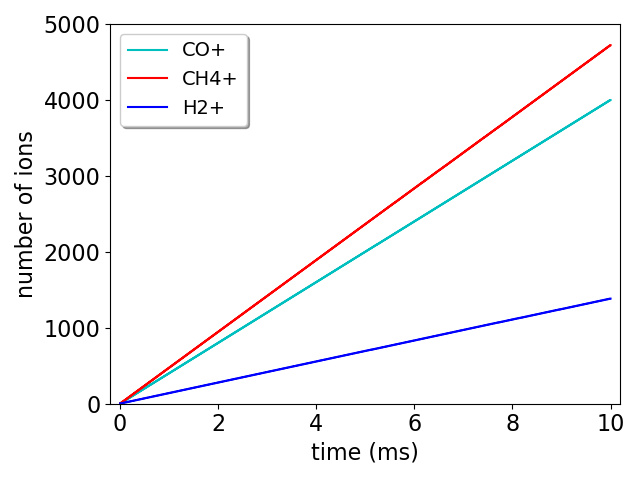}
   \caption{Number of ions of gas A (left) and B (right) in the pipe generated at a length of \unit{4}{\centi\metre} at the multi-purpose electrode of \Berlinpro\ with voltage-configuration~1.}
   \label{bpro_statistics_1000}
\end{figure}
\begin{figure}[]
   \centering
   \includegraphics*[width=0.238\textwidth]{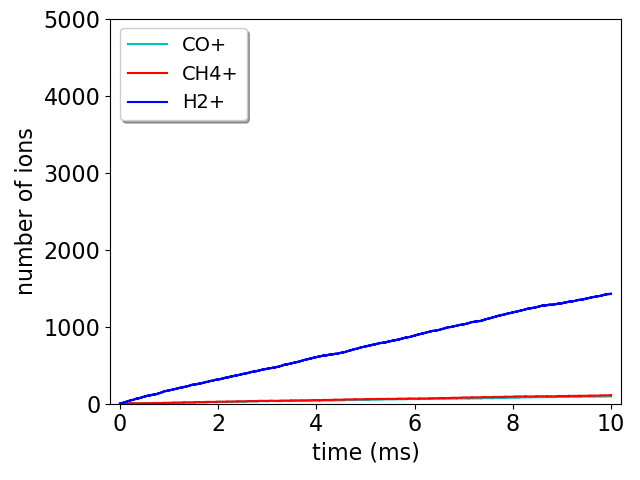}
   \includegraphics*[width=0.238\textwidth]{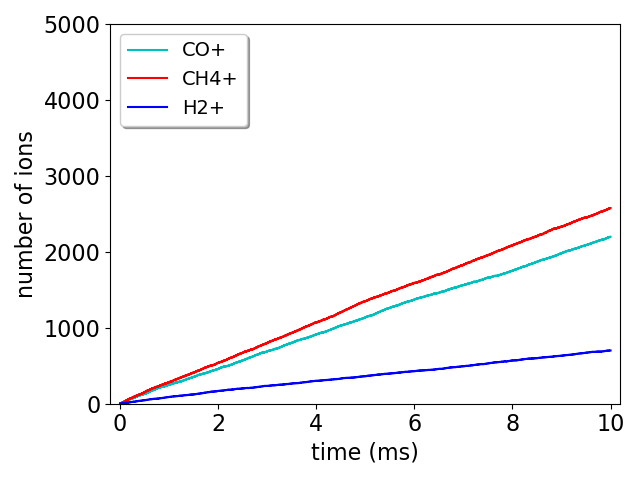}
   \caption{Number of ions of gas A (left) and B (right) in the pipe generated at a length of \unit{4}{\centi\metre} at the multi-purpose electrode of \Berlinpro\ with voltage-configuration~2.}
   \label{bpro_statistics}
\end{figure}
\begin{figure}[]
   \centering
   \includegraphics*[width=0.238\textwidth]{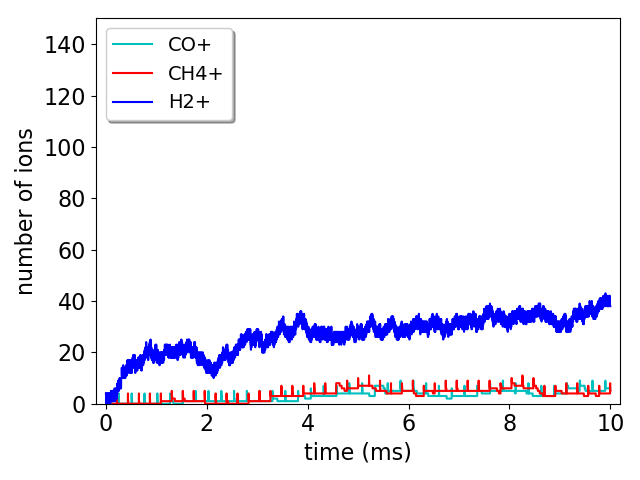}
   \includegraphics*[width=0.238\textwidth]{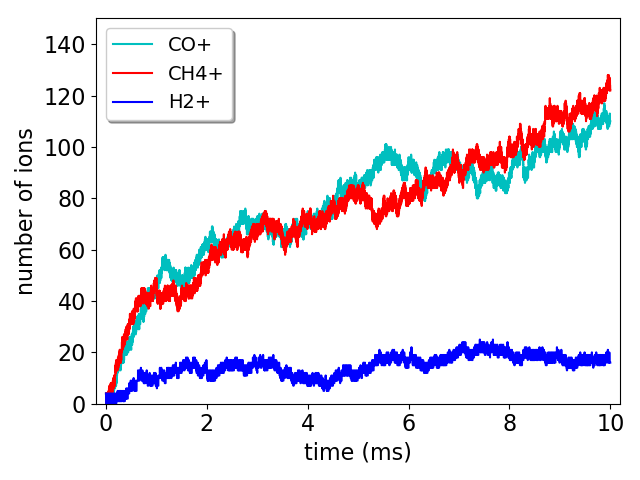}
   \caption{Number of ions of gas A (left) and B (right) in the pipe generated at a length of \unit{4}{\centi\metre} at the multi-purpose electrode of \Berlinpro\  with voltage-configuration~3. }
   \label{bpro_statistics_0750-0750}
\end{figure}
\begin{figure}[]
   \centering
   \includegraphics*[width=0.238\textwidth]{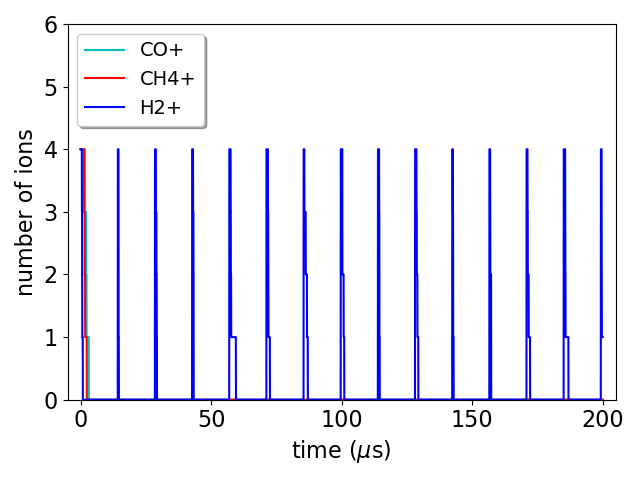}
   \includegraphics*[width=0.238\textwidth]{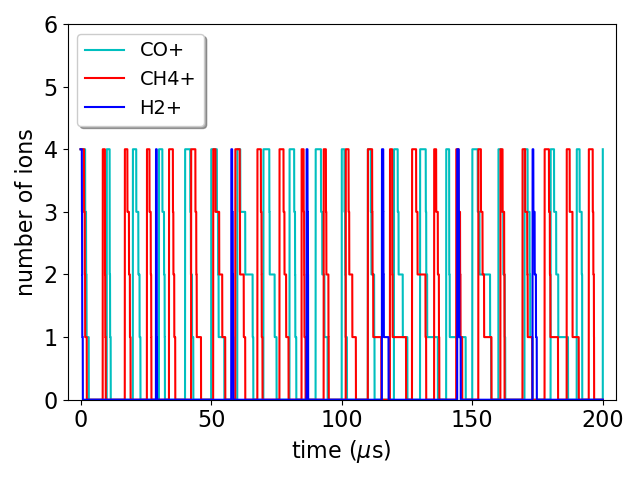}
   \caption{Number of ions of gas A (left) and B (right) in the pipe generated at a length of \unit{4}{\centi\metre} at the multi-purpose electrode of \Berlinpro\  with voltage-configuration~4.}
   \label{bpro_statistics_1000-1000}
\end{figure}

Almost no clearing occurs in the case of voltage-configuration~1 as shown in Fig.~\ref{bpro_statistics_1000}.  The residual potential on axis is obviously not effective. Also the voltage-configuration~2 is not effective. A strong accumulation occurs in this case as shown in Fig.~\ref{bpro_statistics}  
for a simulation time of \unit{10}{\milli\second}. The results of voltage-configuration~3 are represented in Fig.~\ref{bpro_statistics_0750-0750}. It turns out, that almost an equilibrium
between ion generation and clearing occurs for gas mixture A, while for gas mixture B with the same voltage-configuration the accumulation does not stop but the accumulation rate decreases (compared to voltage-configuration~2).
In the case of voltage-configuration~4 all generated ions are almost immediately cleared after generation. 
Fig.~\ref{bpro_statistics_1000-1000}  
represents the first \unit{200}{\micro\second} of the simulations for voltage-configuration~4. Obviously,
the created ions are cleared within a few microseconds.
\par
\begin{figure}[]
   \centering
      \setlength{\fboxsep}{1pt} 
   \setlength{\fboxrule}{0.3pt} 
   \fbox{
   \includegraphics*[width=64mm]{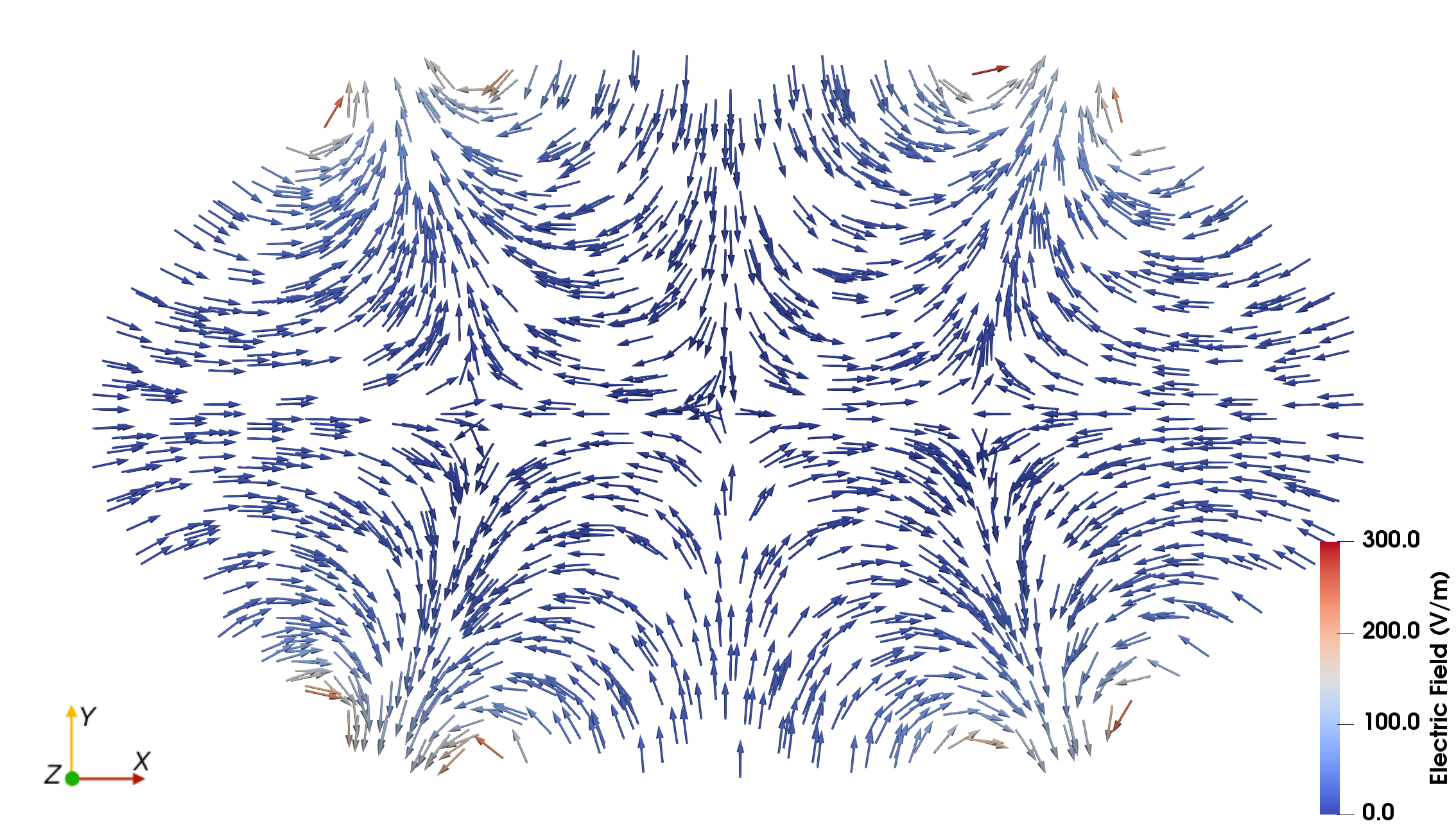}}
   \fbox{
   \includegraphics*[width=64mm]{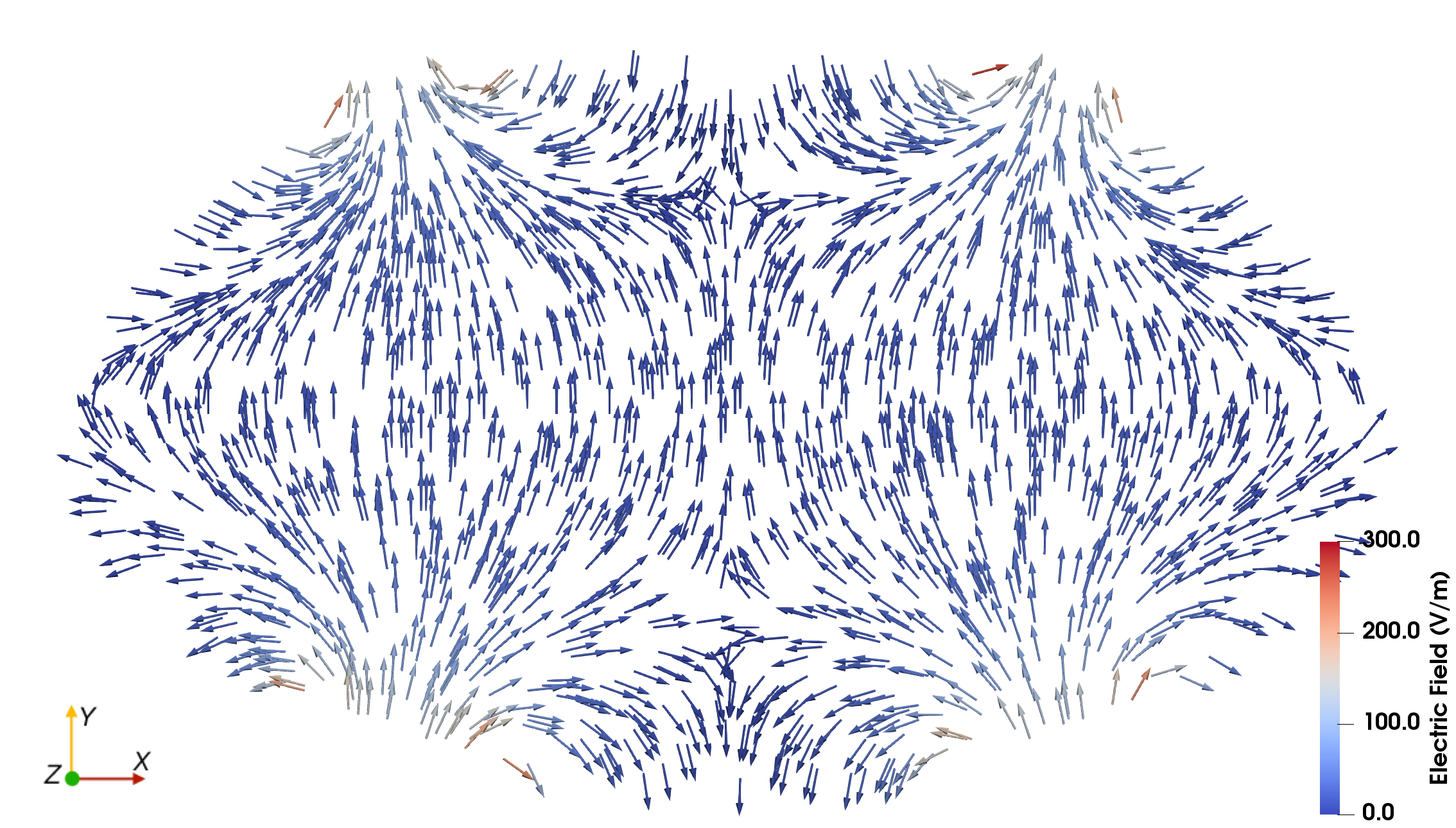}}
   \caption{Electric field of the multi-purpose electrode: voltage-configuration~1 (top) compared to voltage-configuration~4 (bottom); the transversal cross-sections at $z = 0$ are shown. }
   \label{bpro_field_trans}
\end{figure}
\begin{figure}[]
   \centering
      \setlength{\fboxsep}{1pt} 
   \setlength{\fboxrule}{0.3pt} 
   \fbox{
   \includegraphics*[width=64mm]{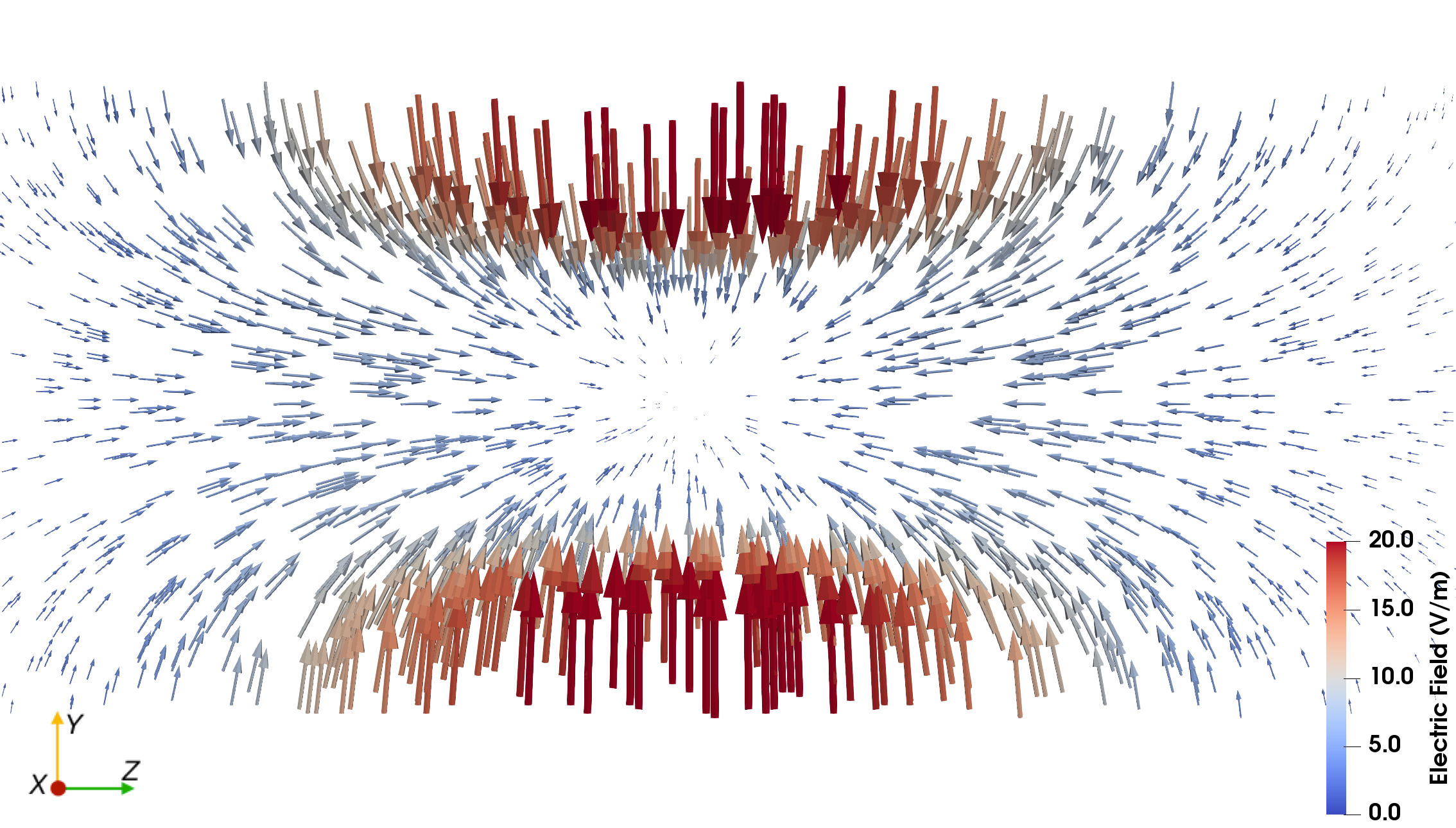}}
   \fbox{
   \includegraphics*[width=64mm]{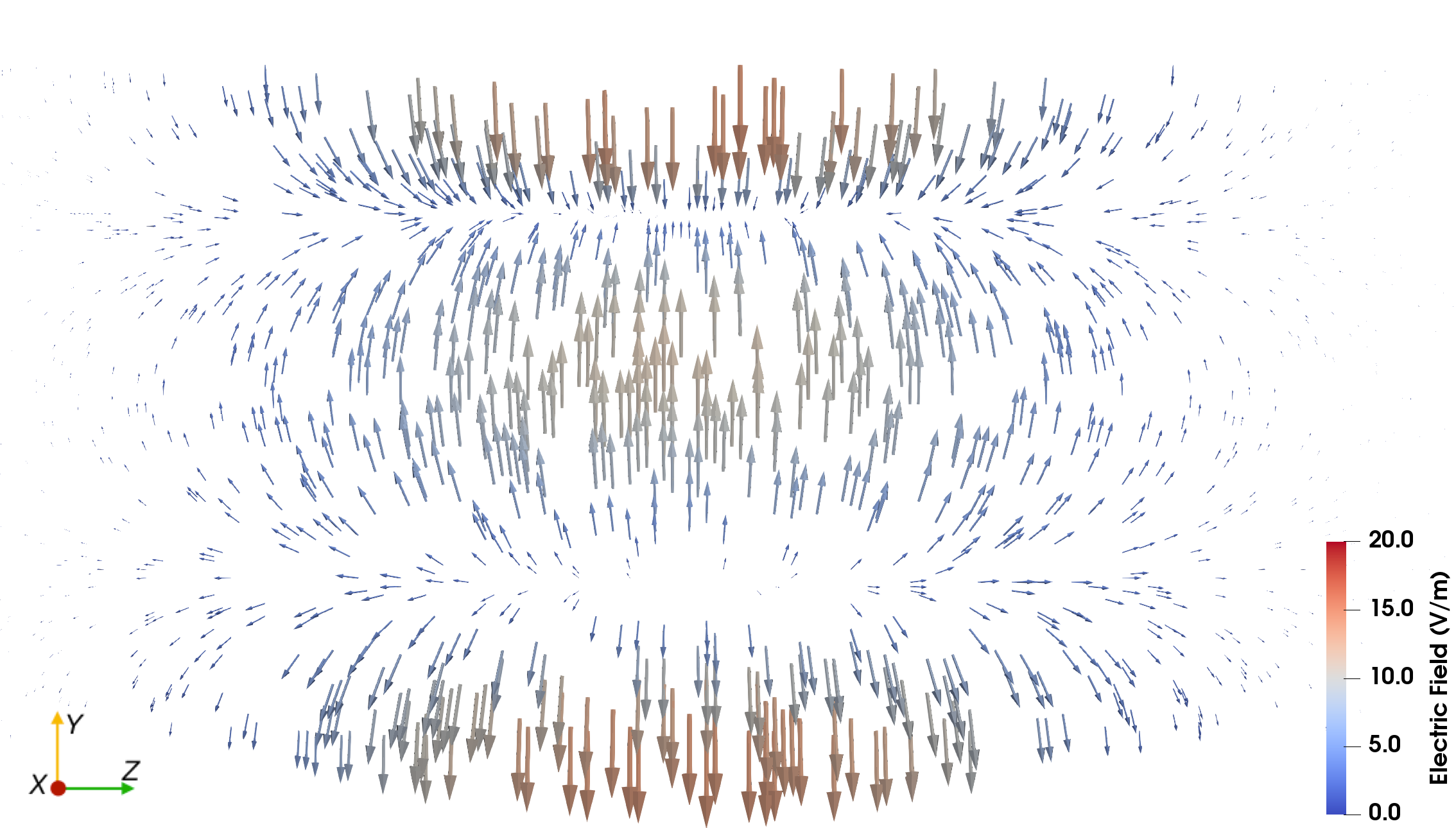}}
   \caption{Electric field of the multi-purpose electrode:  voltage-configuration~1 (top) compared to voltage-configuration~4 (bottom);  the longitudinal cross-sections at $x = 0$ are shown.}
   \label{bpro_field_long}
\end{figure}

Considering the electrical fields within the clearing electrodes for the four voltage-configurations, our simulation results do not surprise. 
Fig.~\ref{bpro_field_trans}~(top) shows the electrical field lines for the voltage-configuration~1.  In the center of the beam pipe the largest part of the field lines run
parallel to the horizontal axis. Hence, they do not guide the ions towards the electrodes directly. In the case of the voltage-configuration~4, shown in Fig.~\ref{bpro_field_trans}~(bottom), the situation is very different. Almost all field lines point towards the electrodes. Thus the ions are guided very efficiently towards the electrodes. Also the longitudinal cross-sections for the two voltage-configurations, shown in Fig.~\ref{bpro_field_long}, confirm this finding.  
\par
\begin{figure}[]
   \centering
   \includegraphics*[width=75mm]{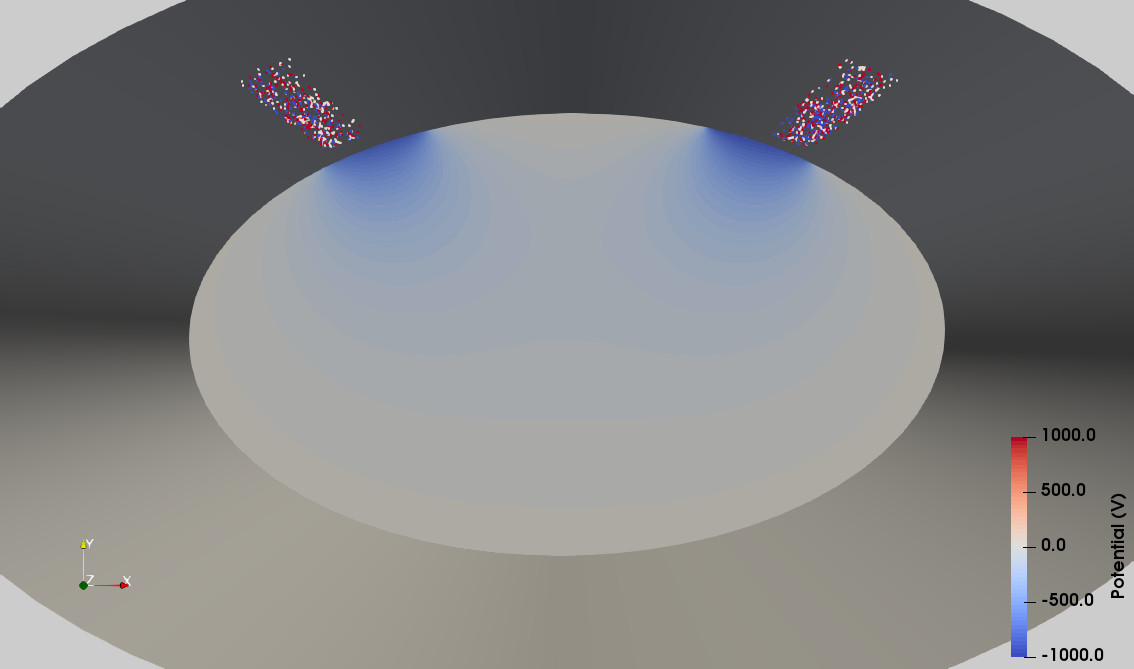}
   \includegraphics*[width=75mm]{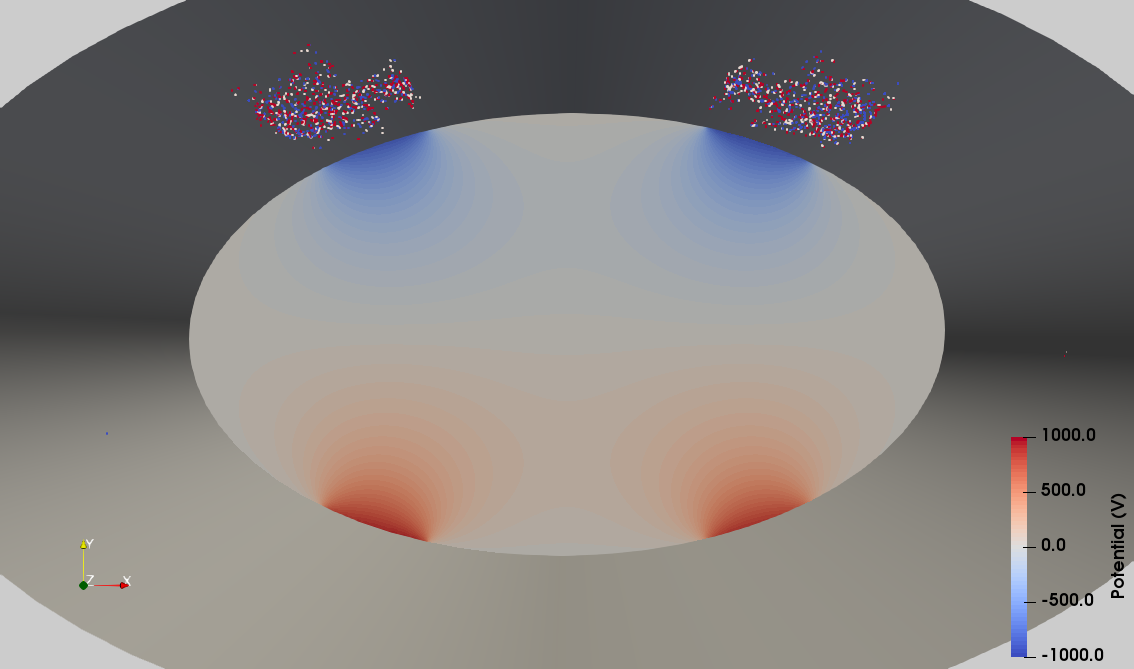}
   \caption{Distribution of the cleared ions (Gas B: \Htwo - blue, \CHfour - white, \CO - red) at the electrodes of voltage-configuration~2 (top) and  voltage-configuration~4 (bottom), respectively.}
   \label{bpro_field1000V}
\end{figure}

In the case of the voltage-configuration~2, the electrical field in the center of the pipe is not strong enough for a significant clearing. Furthermore, the electrical fields of the two upper stripes almost cancel each other in the center of the pipe and the field is stronger outward. The effect of this field configuration is nicely visible in the Fig.~\ref{bpro_field1000V}, where the distributions of cleared ions for voltage-configuration~2 and~3 are depicted. Obviously, the cleared ions hit only the outer-side of the stripes in voltage-configuration~2, whereas in voltage-configuration~3 the whole surface of the stripes is hit. 
\par
Please note that in the generic case of electrode~2 presented in the previous section the distance is much smaller between the two parallel upper and two parallel lower stripes, respectively. Therefore, its field distribution differs from the field distribution for the \Berlinpro\ multi-purpose clearing electrodes. Thus, it is no surprise that the electrode~2 in the generic case shows a good clearing performance, even if all four stripes have the same voltage, while the  \Berlinpro\ multi-purpose clearing electrodes require a strong field gradient in the pipe center for a good clearing performance.     
This field gradient leads also to an undesired deflection of the beam which needs to be compensated by the orbit correction system. We estimate a deflection angle of about \unit{10}{\micro\rad}
in the case of voltage-configuration~4, i.e.\ for the maximum field gradient.
  \par

\par
\par
\par
\par
\par
\section{\label{sec:05}Conclusion}
We have presented numerical studies of the clearing process of continuously generated ions by clearing electrodes. Thereby the passing electron bunches were considered with typical parameters of future high current linacs. Furthermore, detailed performance studies of the multi-purpose clearing electrodes planned for \Berlinpro\ have been discussed. 
\par
For the numerical simulations, the software tool CORMORAN was applied for the ion generation and tracking, where the external fields of the clearing electrodes were pre-computed by means of the Python Poisson Solver of Compaec~e.G. We investigated two gas mixtures and five different designs of electrodes partly with different voltages.  
\par
It could be shown that in the direct volume of the clearing electrodes (\unit{2}{\centi\metre}) all four single-purpose designs have a sufficient performance, when applying clearing voltages of a few \unit{1000}{\volt}. The level of trapped ions within the electrodes remains far below $10\%$ of the total number of generated ions, in spite of the relatively high pressure of \unit{10^{-8}}{\milli\bbar} assumed for the generic case. However, the vicinity of the electrodes can only be cleared by a significant increase in voltage. In this connection, the clearing electrode 2 shows better performance, as it requires the lowest voltage. There is also a hint that the special composition of electrode 2 can be beneficial to the clearing process of heavier ions. 
\par
The application of different voltage-configurations for multi-purpose electrodes showed that a proper clearing performance can only be achieved by biasing the upper and lower stripe-pairs with differently signed voltages. This can be understood by examining the course of the electrical field lines within clearing electrodes for different voltage configurations. Furthermore, the presented simulation studies showed that the multi-purpose electrode allow for a proper clearing with a voltage about \unit{1000}{\volt} in the \Berlinpro\ case. 
\\[2ex]
\begin{acknowledgments}
We would like to thank Volker D\"urr and J\"org Kolbe, both Helmholtz--Zentrum Berlin,  for fruitful discussions. 
\end{acknowledgments}
%
%
%
%
\end{document}